\documentclass[aps,onecolumn,preprint,superscriptaddress,nofootinbib,floats,prd]{revtex4}
\usepackage{amsmath,amssymb,color,mathrsfs, graphicx,verbatim,epsfig, wasysym, multirow}

\def\lsim{\mathrel{\rlap{\lower4pt\hbox{\hskip1pt$\sim$}}
    \raise1pt\hbox{$<$}}}
\def\gsim{\mathrel{\rlap{\lower4pt\hbox{\hskip1pt$\sim$}}
    \raise1pt\hbox{$>$}}}

\newcommand{\be}{\begin{eqnarray}}
\newcommand{\ee}{\end{eqnarray}}

\def\addresses#1#2{\hbox to \hsize{\@tablebox{#1}\hfil\@tablebox{#2}}}
\def\@tablebox#1{\vtop{\hsize=5in \begin{flushleft} #1 \end{flushleft}}}

\def\beq{\begin{equation}}
\def\eeq{\end{equation}}
\def\bit{\begin{itemize}}
\def\eit{\end{itemize}}

\def\beqarray{\begin{eqnarray}}
\def\eeqarray{\end{eqnarray}}

\def\met{$\displaystyle{\not}E_T$}

\begin{document}

\begin{titlepage}

\thispagestyle{empty}

\begin{flushright}UMD-PP-10-018
\end{flushright}
\vspace{0.2cm}

\begin{center}

\vskip .5cm

{\Large \bf  Ditau-Jet Tagging and Boosted Higgses from a \\ \vspace{0.3cm} 
Multi-TeV Resonance }
\vskip 1.0cm
{\large Andrey Katz,$^1$ Minho Son,$^2$ and Brock Tweedie$^{3,4}$}
\vskip 0.4cm
{\it $^1$Department of Physics, University of Maryland, College Park, MD 20742} \\
{\it $^2$Department of Physics, Yale University, New Haven, CT 06511} \\
{\it $^3$Department of Physics and Astronomy, Johns Hopkins University, Baltimore, MD 21218} \\
{\it $^4$Physics Department, Boston University, Boston, MA 02215}
\vskip 1.2cm

\end{center}

\noindent
New TeV-scale physics processes at the LHC can produce Higgs bosons with substantive transverse Lorentz boost, such that the Higgs's decay products are nominally contained in a single jet.  In the case of a light Higgs decaying predominantly to $b\bar b$, previous studies have shown that these Higgs-jets can be identified by capitalizing on jet substructure techniques.  In this work, we explore the possibility of also utilizing the subdominant but very distinctive decay $h\to\tau^+\tau^-$.  To this end, we introduce the concept of a ``ditau-jet,'' or a jet consisting of two semi-collinear taus where one or both decay hadronically.  We perform simple modifications to ordinary tau tagging methods to account for this configuration, and estimate tag rates of $O$(50\%) and QCD mistag rates of $O(10^{-4}-10^{-3})$ for $p_T \sim$~TeV, even in the presence of pileup.  We further demonstrate the feasibility of reconstructing the ditau invariant mass by using traditional \met\ projection techniques.  Given these tools, we estimate the sensitivity of the LHC for discovery of a multi-TeV $Z'$ decaying to $Zh$, utilizing both leptonic and hadronic $Z$ decay channels.  The leptonic $Z$ channel is limited due to low statistics, but the hadronic $Z$ channel is potentially competitive with other searches.

\end{titlepage}

\setcounter{page}{1}

\section{Introduction}
\label{sec:intro}


As the first collider capable of probing multi-TeV energy scales, the LHC will face a unique set of challenges in reconstructing 
some of its most extreme events.  Particles which have traditionally been considered heavy can be produced with energies far in 
excess of their mass, leading us to consider novel classifications such as ``top-jets,'' ``$W$-jets,'' ``$Z$-jets,'' and ``Higgs-jets.''
While these objects can look uncomfortably similar to ordinary QCD jets, in the past several years a variety of techniques have 
been developed to robustly discriminate them~\cite{Seymour:1993mx,Butterworth:2002tt,Butterworth:2008iy,Brooijmans:2008zz,Thaler:2008ju,Kaplan:2008ie,Plehn:2009rk,Plehn:2010st,Rehermann:2010vq,Butterworth:2009qa,Kribs:2009yh,Almeida:2008yp,Almeida:2010pa,Ellis:2009su,Krohn:2009th,Thaler:2010tr}.

In a previous paper~\cite{Katz:2010mr}, we made a detailed study of hadronic $W$-, $Z$-, and Higgs-jets in the context of the search for a multi-TeV 
$Z'$, which commonly arises in composite Higgs and 5D models as well as in simple extended gauge sectors (see, e.g.,~\cite{Agashe:2003zs,ArkaniHamed:2001nc,Schmaltz:2005ky,Langacker:2008yv} and others in~\cite{Katz:2010mr}).  Our previous study was restricted to the case where the boson-jet 
consists of quark-antiquark, which is a shared option in the case that the Higgs is light and its decays 
dominated by $b\bar b$.  However, since the mass and branching fractions of the Higgs are as yet unknown, it is important to 
consider a broader set of options.

In this paper, we explore the possibility of utilizing Higgs-jets generated by the decay $h\to\tau^+\tau^-$.  We call these 
``ditau-jets.''  For a standard Higgs near 120 GeV, which is favored by electroweak precision measurements, this mode represents 
about 7\% of the total decay rate and quickly drops as the mass is raised.  While this means that the $\tau^+\tau^-$ mode is always 
subleading, several considerations compel us to take a closer look at it.  First, from the perspective of discovering multi-TeV 
physics that generates Higgs-jets, establishing additional search channels increases our chances of spotting something new, and 
increases our confidence if we do.  Second, from the perspective of testing Higgs physics, obtaining a measurement of this rate with 
respect to $b\bar b$ can give us a measure of the Higgs couplings, and test whether the ``Higgs'' observed in a new physics process 
is indeed the one observed at lower energies.  Finally, from the perspective of signal discrimination, a high-$p_T$ ditau-jet is 
in principle much easier to spot than a high-$p_T$ $b\bar b$-jet, as we will see.

Of course, these observations are all strictly theoretical if the ditau-jet production rate is genuinely too small to see.  In order 
to make this Higgs decay mode truly competitive without running the LHC ten times longer, there needs to be a tradeoff in branching 
fraction with another part of the event.  The process $Z'\to Zh$ provides a ready example.  Here, one of the main discovery 
channels with a light Higgs is $Zh\to(l^+l^-)(b\bar b)$~\cite{Atl:LH,Agashe:2007ki,Katz:2010mr}, with branching fraction penalties of roughly 7\% on the $Z$ and 70\% on 
the Higgs.  We can instead consider the channel $Zh\to(q\bar q)(\tau^+\tau^-)$, with respective penalties of about 70\% and 7\%.  
In order for this to be a worthwhile tradeoff, we must demonstrate that a ditau-jet can be very cleanly identified, as was the 
case for the dileptonic $Z$ in the first channel.  If this can be accomplished with high efficiency and high purity, then the 
now-hadronic $Z$ on the other side of the event can also be tagged using jet substructure (analogous to the $b\bar b$ Higgs-jet), and we in principle 
obtain $Z'$ sensitivity at least as good as what was obtained in the untagged analysis of $Zh\to(l^+l^-)(b\bar b)$ 
in~\cite{Katz:2010mr}.  The key question is:  How well can we discriminate ditau-jets from QCD?

There are two basic facts about taus that should allow this discrimination to be accomplished quite well, at least in principle.  
The first is that almost 60\% of ditau decays contain at least one lepton.  The second is that hadronic tau decays consist of 
extremely tight cones of activity (mostly charged and neutral pions), with opening angles of $O(m_\tau/p_{T\tau})$.  
At progressively higher energies, these cones become progressively smaller.  QCD jets, by contrast, are color-connected to the rest of the event and 
tend to radiate at all angles.  Highly energetic hadronic taus therefore allow us an extreme form of the kind of color-discrimination explored 
in~\cite{Gallicchio:2010sw,Falkowski:2010hi}.  This improves as the energy is raised, suggesting that the hadronic taus produced in 
TeV-scale processes should be particularly clean.

Nonetheless, the case of two nearby high-$p_T$ taus has not been studied in detail before.  The decays that proceed dileptonically 
should be straightforward to find, but cannot uniquely be ascribed to taus, and are rather rare (12\% $BR$).  The remaining decays 
contain at least one hadronic tau in close proximity to either a lepton or another hadronic tau.  These configurations can fail 
normal tau and/or lepton identification criteria.  To deal with this possibility, we propose simple extensions of tau and lepton 
isolation criteria such that two objects can be ``mutually isolated,'' i.e.\ individually isolated from tracker and ECAL activity 
not caused by the other object.  This forms the basis of a ditau-tag, which can be applied in both semi-leptonic and fully 
hadronic ditau decays, and extends to TeV-scale energies and very small tau separations.

We first study mutual isolation in the context of 2 and 3 TeV $Z'\to Zh\to(l^+l^-)(\tau^+\tau^-)$, finding ditau-tagging efficiencies 
of order 50\% with QCD rejection rates at the $10^{-4}-10^{-3}$ level.  We also demonstrate that reconstruction 
of the Higgs mass peak should be possible, even given the presence of two neutrinos and nearly-collinear geometry.  We proceed to estimate the $Z'$ 
discovery reach in this channel and in the $(q\bar q)(\tau^+\tau^-)$ channel, demonstrating that the latter is a possible 
competitor to the diboson final-state channels that we studied in~\cite{Katz:2010mr}.  

The paper is organized as follows.  In the next section we review tracker-based hadronic tau-tag techniques and introduce 
our ditau-tag.  In the third section we categorize the performance of the tag and ditau mass reconstruction, and estimate $Z'$ 
discovery potential.  We conclude in section~\ref{sec:outlook}.

\section{Tagging Boosted Taus and Ditaus}
\label{sec:tau_substructure}


In this section, we outline techniques for tagging hadronic taus at high $p_T$.  We start by reviewing some of the techniques that will be applied to single hadronic taus at the LHC.  We then move on the case relevant for boosted Higgs decays, namely tagging two taus in close proximity.  This includes cases where one or both of the taus decays hadronically.  In addition, 12\% of ditaus will decay dileptonically, but for these we do not pursue any special techniques.

\subsection{Review of hadronic tau tag}\label{singletautag}

In some sense, hadronically decaying taus can be treated as ``fat'' leptons.  For electrons and muons, a track and/or small cluster of electromagnetic activity is first identified, and then demanded to be isolated from any additional activity out to a certain $\Delta R$ range and up to a certain energy threshold.  Hadronic taus are not so well-localized in the detectors, since they decay before reaching the inner tracker.  However, the multiplicity of decay products is always fairly small, and their $\Delta R$ separation is characterized by the ratio $m_\tau/p_{T\tau}$, which becomes quite small at even modestly large energies.  The practical way to identify a tau, then, is to define a small region of hadronic and electromagnetic activity in the detector, and insist that it is isolated from additional activity.

The first step is to define the region of interest.  This is done using either a hard track or a hard calorimeter cell as a seed, around which further activity is sought out (see, e.g.~\cite{TauCalAtlas,TauTrackAtlas}).  For typical analysis with taus, the choice of track seed or calorimeter seed is not so crucial, and algorithms built on each of these have similar performance.  However, as we will ultimately be concentrating on the TeV-scale momentum regime with {\it two} nearby taus, we choose to focus on a track-seeded approach, which should work more robustly when dealing with small angular scales.  It would be interesting to further pursue the extension of calorimeter-seeded approaches to ditau-jets, but we do not do so here.

Let us now review the CMS particle-flow algorithm for tau-tagging, as described in~\cite{CMStau}.  After reconstructing jets with a $R=0.5$ cone algorithm, an individual jet is investigated for constituent hadronic tracks.  A track will qualify as a seed for hadronic tau-tagging if
\begin{itemize}
 \item it is the hardest track in the jet
 \item it has $p_T > 5$ GeV
 \item it deviates by no more than $\Delta R= 0.1$ from the jet vector.
\end{itemize}
If one finds such a seed, at the next step an \emph{inner cone} is defined around the seeding track, where all of the 
products of the hadronic tau decay are expected to be found.  For sufficiently high $p_T$, the radius of this 
cone is taken to be 0.07.\footnote{The CMS particle-flow tau-tag actually takes advantage of an inner cone radius that shrinks with $p_T$, tracing the expected physical shrinking of tau decay angles.  However, the minimum cone size allowed is 0.07, which is applied to taus with $p_T > 70$ GeV.}

\begin{figure}[t]
\centering
\includegraphics[width=2.3in]{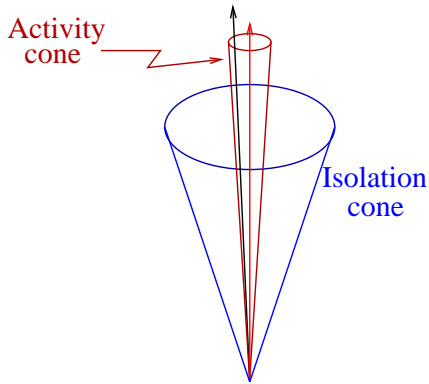}
\caption{{\it Illustration of tracker-seeded hadronic tau tag.}}
\label{fig:A1}
\end{figure}

The decay products of the tau should be inside this inner cone, and almost nothing should be outside.  To enforce isolation, one further defines an isolation annulus, namely an outer cone of $R_{\rm outer}=0.45$ around the leading prong which omits the contents of the inner cone (Fig.~\ref{fig:A1}).  Within the annulus, there should be no photons with $p_T > 1.5 \ {\rm GeV}$ and no tracks with $p_T > 1\ {\rm GeV}$.  (The hadronic calorimeter, which is the least spatially-precise subdetector, is not utilized for this isolation.)  Inner cones of activity which are constructed and isolated in this manner are considered tau-tagged.  As shown in~\cite{CMStau}, the efficiency of this algorithm is about 80\% in the central region of the detector, and the QCD fake rate is at the few percent scale.  The fake rate declines as $p_T$ is increased, while the tag rate plateaus.  This makes sense physically, since the tau decay products become more collimated whereas QCD jets continue to radiate at all angles.

\subsection{Ditau tag} \label{algorithms}

Let us now imagine a ditau-jet, such as would be produced in the decay of a highly energetic light (120 GeV) Higgs from a TeV-scale process.  We restrict ourselves to cases where at least one of the two taus decays hadronically.  If the $p_T$ of the Higgs is 1 TeV, then the $\Delta R$ separation between the two taus will be roughly 0.3, which would lead to failure of the single tau-tagging described above.  However, the pattern of activity will still look highly unusual compared to an ordinary QCD jet, especially at the energies under consideration.

\begin{figure}[t]
\centering
\includegraphics[width=2.3in]{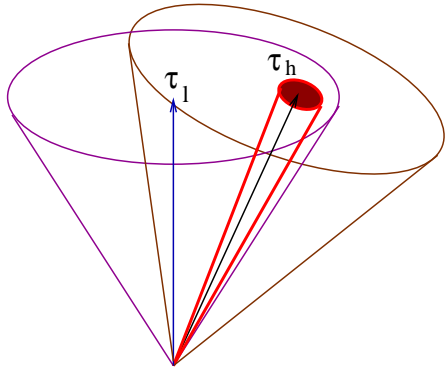}
\caption{{\it Illustration of mutual isolation cones and inner activity cone for a semileptonic ditau.}}
\label{fig:ditau1}
\end{figure}

The simplest and most robust case is semileptonic, which accounts for about 45\% of ditau decays.  The lepton can nominally fail isolation because of the nearby hadronic tau, and vice versa.  As a remedy, we propose the concept of \emph{mutual isolation}.  We call the lepton and $\tau$-jet mutually isolated if the jet passes the criteria for a hadronic tau when the lepton is removed from consideration, and at the same time the lepton passes some form of isolation criterion with the tau inner cone removed.  Step-by-step, this runs as follows:
\begin{itemize}
 \item find a lepton with $p_T > 30$ GeV which fails simple isolaton criteria using an $R=0.4$ cone
 \item find the hardest hadronic track inside this cone with $p_T > 20$ GeV, and bail out if none is found
 \item draw a small tau-candidate cone of radius $R_{\rm inner} = 0.07$ around this seed track
 \item check whether the lepton passes a weakened isolation criterion using a surrounding $R=0.4$ cone with this small cone deleted, accounting only for tracker and electromagnetic energy: $\frac{p_T(l)}{p_T(l)+p_T({\rm cone})} > 0.9$
 \item if the lepton passes, independently cluster all additional activity in the vicinity (we use $R_{\rm jet}=0.4$ Cambridge/Aachen(C/A)~\cite{Dokshitzer:1997in,Wobisch:1998wt})
 \item apply the single tau-tag the resulting jets, ignoring the lepton for purposes of annulus isolation
 \end{itemize}
Schematically this algorithm is illustrated in Fig.~\ref{fig:ditau1}.  We will categorize its performance in the context of a resonance search the next section.

\begin{figure}[t]
\centering
\includegraphics[width=2.3in]{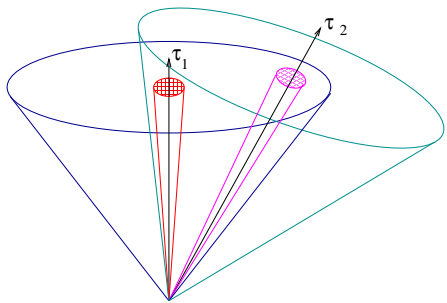}
\caption{{\it Illustration of mutual isolation cones and inner activity cones for an all-hadronic ditau.}}
\label{fig:ditau2}
\end{figure}

It is not difficult to extend this mutual isolation technique to the fully hadronic case, which accounts for about 42\% of ditau decays.  Now, we start with a candidate jet of $R_{\rm jet}=0.4$ (clustered with C/A, though the choice of algorithm is not crucial) and search inside of it for {\it two} tracks which can seed independent taus:
\begin{itemize}
\item  from amongst the jet constituents, use the hardest charged hadron as the first seed, and then look for the hardest track $\Delta R > 0.08$ away from this for the second seed (again, each should be above 20 GeV in $p_T$)
 \item draw a small cone around each seed, picking the inner radius depending on their separation:  
$ R_{\rm inner} = \Delta R_{\rm seeds}/2$ if the distance between the seeds is smaller than 0.14, or $R_{\rm inner} = 0.07$ if the distance between the seeds is bigger\footnote{The occurrence of $\Delta R_{\rm seeds} < 0.14$ is not very common in our analysis below.  In particular, accepting these configurations allows us to recapture about 5\% of the dihadronic $h\to\tau^+\tau^-$ from 3 TeV resonance decay.  However, a shrinking inner cone can be important for even higher-energy searches, or for ditau produced from boosted particles lighter than 120 GeV.}
 \item draw an isolation annulus ($R_{\rm outer} = 0.4$) around each inner cone, with the other inner cone deleted, and check that each tau candidate passes tau isolation criteria
 \item if the candidates are mutually isolated, then use the original jet to define the ditau four-vector, and the seed directions to define the individual tau trajectories.
\end{itemize}
This procedure is sketched in Fig.~\ref{fig:ditau2}.  We also categorize its performance in the next section.

There are potentially several basic limitations to these procedures.  An important one is the ability of the ATLAS and CMS detectors to cleanly identify tracks and photons at such high energies and small angular separations.  In particular, tracking may fail and nearby ECAL cells will share energy to some extent.  However, we do not expect that these will pose insurmountable problems.  Ideally, a genuine hadronic tau consists of only one or three (or very rarely five) tracks, implying a very modest density of hits in the tracker.  This low track (or hit) multiplicity could even be used to strengthen the discrimination against QCD.  The caveat is that these tracks are almost always accompanied by photons from $\pi^0$s, which can convert in the tracker (especially at high energy), and increase the activity.  We have no way to model this effect in our study, but simply assume that nearby photon conversions do not dramatically degrade tracking quality as we move from 100 GeV taus to 1000 GeV taus.  In the ECAL, energy will spread laterally, and we may worry that some fraction will enter the isolation annulus.  But the characteristic distance scale of this leakage is fairly energy-independent and is of the same order as the cell size.  Our inner cones are intentionally taken to be 3-4 cells in radius, and should contain the vast majority of the electromagnetic activity, especially since the photons are becoming more collinear at higher energies.  In the studies below, we will nonetheless incorporate some of the detector effects by using a primitive detector model consisting of separate tracker, ECAL, and HCAL elements.  This is mainly relevant for obtaining quasi-realistic estimates of energy and mass resolutions.

The other major worry is pileup, which will introduce a spray of uncorrelated particles into the isolation annuli.  We find that if we simply accept these additional particles and run the ditau tags using the isolation criteria of subsection~\ref{singletautag}, the signal rates drop by a factor of order 3 due to isolation failures.  However, pileup consists of min-bias events produced at distinct vertices.  This means that pileup tracks can in principle be removed by tracing them back to the ``wrong'' vertex.  Assuming that this can be done with high efficiency, we are still left with irreducible pileup neutrals, but these do not tend to be very energetic.  We will see below that raising the annulus photon threshold from 1.5 GeV to 2.5 GeV is quite adequate.  The tag rates end up only 10\% lower, relatively speaking, than what is obtained without pileup using the original threshold, and the fake rates remain acceptable.  It is also worth noting that even if pileup tracks cannot be efficiently removed by vertexing, using a slightly higher track $p_T$ threshold would likely have similar impact.

\section{Heavy Resonance Searches Utilizing Ditau-Jet Tagging}
\label{sec:ditau}


In this section, we apply the ditau-tagging of section~\ref{sec:tau_substructure} to the case of a multi-TeV $Z'$ decaying to $Zh$ with a 120 GeV Standard Model-like Higgs boson.  We begin with the case of $Zh\to(l^+l^-)(\tau^+\tau^-)$.  Given that the branching fraction $h \to \tau^+ \tau^-$ is only 7\%, we do not expect this mode to be competitive with the much more prevalent $Zh\to(l^+l^-)(b\bar b)$.  Nonetheless, this serves as a clean context in which to check the performance of the ditau tag and to explore the possibility of Higgs mass reconstruction.  Next, we turn to the more challenging mode $Zh\to(q\bar q)(\tau^+\tau^-)$.  The backgrounds here are potentially severe, but the starting rate now competes with $(l^+l^-)(b\bar b)$.  We will see that ditau-tagging and Higgs mass reconstruction can make the mode viable.\footnote{There are other signal topologies which we might consider.  For example, we could also gain some rate above $Z\to l^+l^-$ by using  $Z\to\nu\bar\nu$.  This can in principle be fairly clean, but the final state always contains sets of neutrinos which are approximately back-to-back, potentially seriously complicating $Z'$ and Higgs reconstruction.  Also, the mode $Zh\to(\tau^+\tau^-)(b\bar b)$ is superficially nearly equivalent to $(q\bar q)(\tau^+\tau^-)$, and could be mined for signal.  However, the starting rate is about a factor of two smaller, and will be cut down by our $Z$ and Higgs mass cuts.  Combining in a dedicated analysis of this mode might also be useful to explore.}

The $Z'$ samples for these studies are generated with 2~$\to$~4 matrix elements in {\tt MadGraph/ MadEvent v4.4.32}~\cite{Alwall:2007st} and showered/hadronized in {\tt PYTHIA} (plugin version {\tt 2.1.3}, taus decayed with {\tt TAUOLA}).  The $Z'$ width is taken to be 3\%, which is smaller than the instrumental width and the signal windows that we use below.  (Our estimates should work reasonably well for resonances up to about 15\% natural width.)  Our background samples are mainly generated with {\tt PYTHIA v6.4.11}~\cite{pythiamanual} with default settings.  This automatically includes prompt and radiative heavy flavor contributions, which can contaminate our semileptonic ditau analysis.  The $W$-strahlung background is generated with {\tt MadGraph}.  All processes are evaluated at leading-order for a 14 TeV LHC.  In addition to the hard collision, each event is superimposed with a pileup of min-bias collisions.  The number of extra collisions is drawn from a Poisson distribution with a mean of 20.  Charged pileup particles are subsequently removed, under the assumption that these could mostly be traced back to distinct vertices.

The particle-level events are further processed through a primitive detector model, corresponding to a kind of idealized particle-flow reconstruction.  Charged particle four-vectors are left untouched.  Photons are passed into a perfect energy sampling ECAL with granularity $\Delta\eta\times\Delta\phi = 0.02\times0.02$.  Neutral hadrons are passed into a perfect energy sampling HCAL with granularity $\Delta\eta\times\Delta\phi = 0.1\times0.1$.  The calorimeter cells are then mapped into massless four-vectors.  We cluster these detector-level particles into jets using {\tt FastJet 2.4.1}~\cite{Cacciari:2005hq}. 

Leptons and jets (or subjets) are subsequently smeared.  Electrons are smeared by 2\%, and muons by $(5\%)\sqrt{E/{\rm TeV}}$.  Jets and subjets are smeared by the resolution quoted in the CMS TDR for $R=0.5$ cone jets~\cite{CMSTDR}:  $\Delta E/E = 5.6/E \oplus 1.25/\sqrt{E} \oplus 0.033$, with $\oplus$ indicating quadrature sum and $E$ measured in GeV.

We estimate discovery reach for 2 and 3 TeV $Z'$ by using simple counting.  For each sample, we construct a box in the $(m_{h}^{\rm reco},m_{Z'}^{\rm reco})$ plane, centered on the signal.  The box is not optimized, other than coarsely by eye.  We claim that discovery is possible if two criteria are met:  $5\sigma$ significance according to Poisson statistics,\footnote{More specifically, we allow $N_s$ for which the integrated probability for background to fluctuate to $N_s$ events or larger is less than $3\times10^{-7}$.} and $N_S > 10$.

\subsection{Leptonic $Z$ and ditau tag efficiency}\label{Zllhtt}

In this analysis, we first identify isolated leptons.  Within a cone of size 0.4, the lepton should account for 90\% of the energy, neglecting photons and any other leptons.  We then seek out an opposite-sign same-flavor pair with $m_{ll} = [75,105]$ GeV to serve as the $Z$ candidate.  We further seek out leptons which fail basic isolation but which are mutually isolated from hadronic taus.  We then cluster the remaining particles in the event into $R=0.4$ C/A jets and check them for either single hadronic tau tag or hadronic ditau tag as in section~\ref{sec:tau_substructure}.  (For the single tag, we use an outer annulus radius of 0.4, in contrast to the CMS choice of 0.45.)  We require events to have a good $Z$ candidate and {\it exactly} two tau candidates (hadronic taus and/or leptons).\footnote{In the fairly rare cases with two sets of opposite-sign same-flavor lepton pairs in the $Z$ mass window, we take the one that comes closest to the $Z$ pole mass as the $Z$ candidate, and the other two as the ditau candidate.  This subjects us to some $ZZ$ background which we do not account for, but could easily be removed with a double-$Z$ veto.}  The vector sum of the visible tau candidate activity should have $|\eta| < 1.5$, placing the taus in the region where the ECAL granularity is highest in the real detectors.

Next, we reconstruct the ditau invariant mass and the $Z'$ mass.  A clear obstacle here is the presence of {\it at least} two neutrinos in the final state from the tau decays.  Nonetheless, we know that these neutrinos are very well-aligned with the visible tau decay products.  Using this fact, and given the measured components of missing transverse energy (\met), it is possible to determine the neutrino three-vector associated to each tau and to fully reconstruct the Higgs.  This is a standard method for light Higgs searches in vector boson fusion.  (Details can be found, for example, in~\cite{Ellis:1987xu,Rainwater:1998kj,Plehn:1999xi,Aad:2009wy}.)  Reconstruction of the $Z'$ is even more straightforward, since all of the neutrinos are fairly collinear from the more global perspective of the entire event.  This reconstruction is not very sensitive to the detailed energy sharing between neutrinos from the different taus.

\begin{figure}[t]
\begin{center}
\epsfxsize=0.47\textwidth\epsfbox{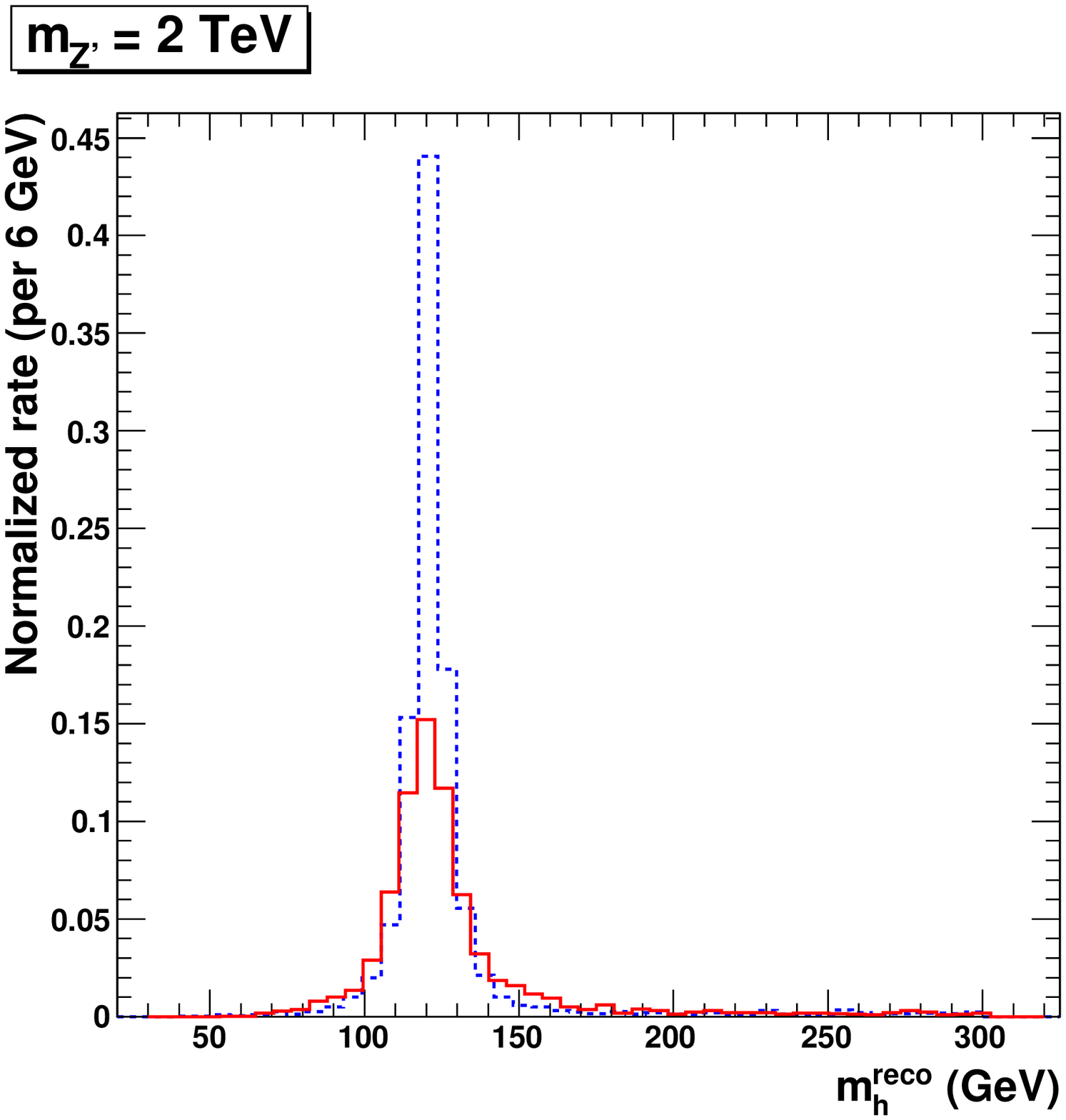}
\epsfxsize=0.47\textwidth\epsfbox{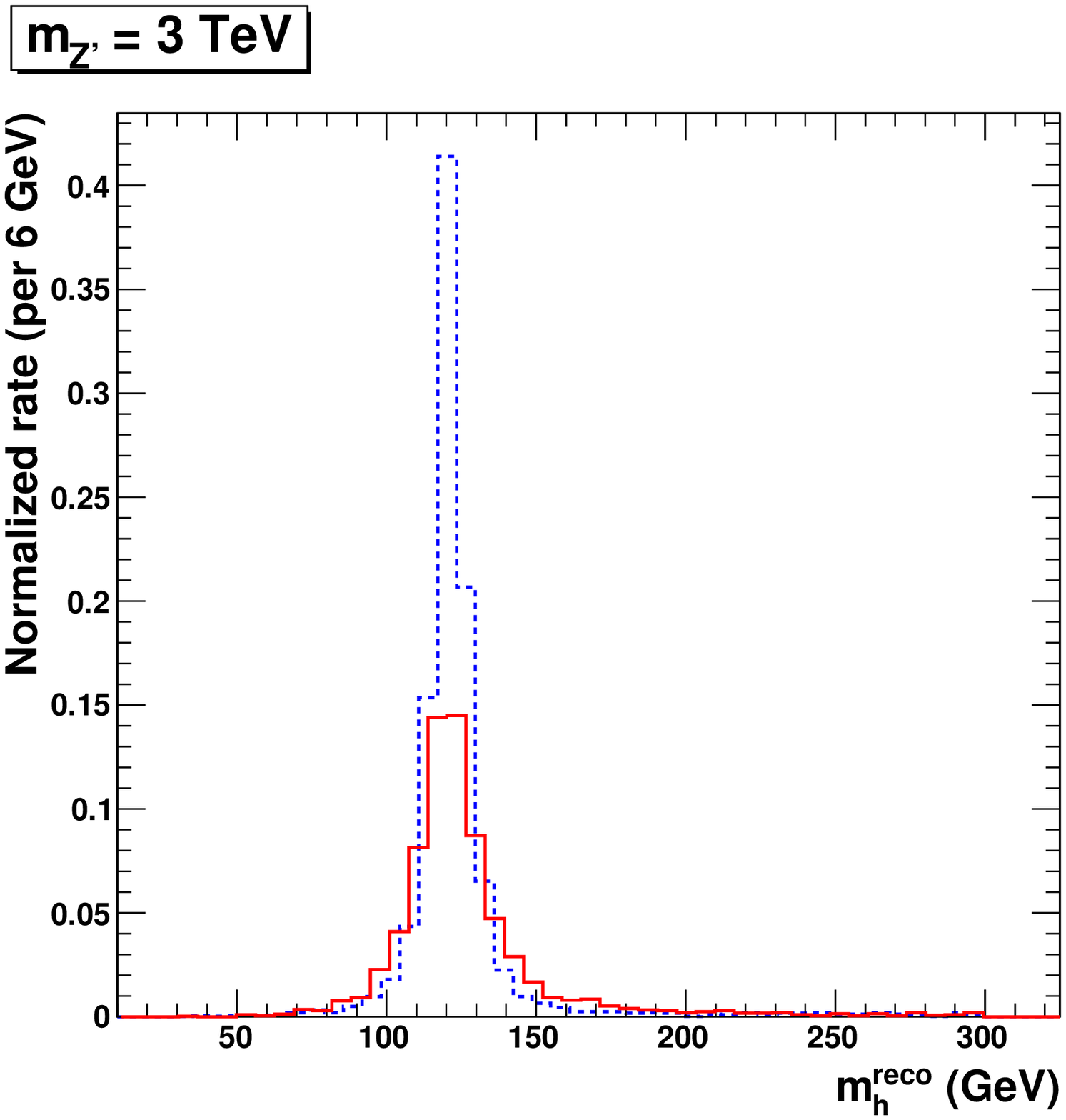}
\caption{{\it Reconstructed Higgs invariant mass from 2 TeV (left) and 3 TeV (right) $Z'$ decays.  The solid red line shows Higgs 
reconstruction using our default modeling of the \met\ measurement, and the dashed blue line shows Higgs 
reconstruction if perfect \met\ is used.}}
\label{fig:ditau3}
\end{center}
\end{figure}

The quality of the Higgs mass reconstruction will depend sensitively on the quality of the missing energy measurement, especially its direction.  While we do not have the tools to model this accurately, we will make the case that this reconstruction should be possible.  We start by defining the \met\ vector to be the transverse vector that balances the sum of the reconstructed $Z$ and the visible products of the two taus.  We re-emphasize that these have been nominally smeared according to expectations at the LHC, and note that this construction of \met\ somewhat pessimistically subjects us to fluctuations from ISR and other uncorrelated event activity.  

\begin{figure}[t]
\begin{center} 
\epsfxsize=0.47\textwidth\epsfbox{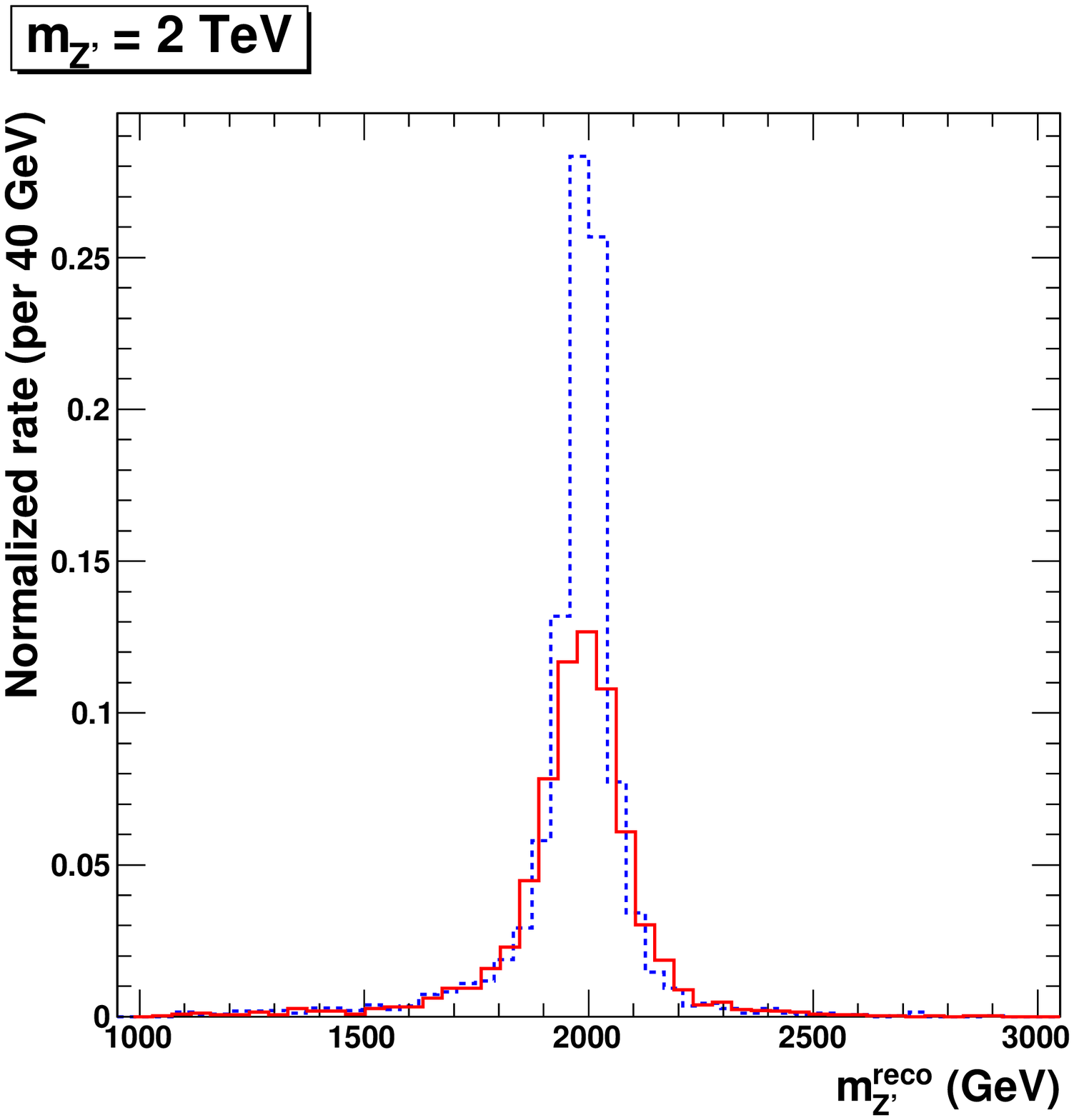}
\epsfxsize=0.47\textwidth\epsfbox{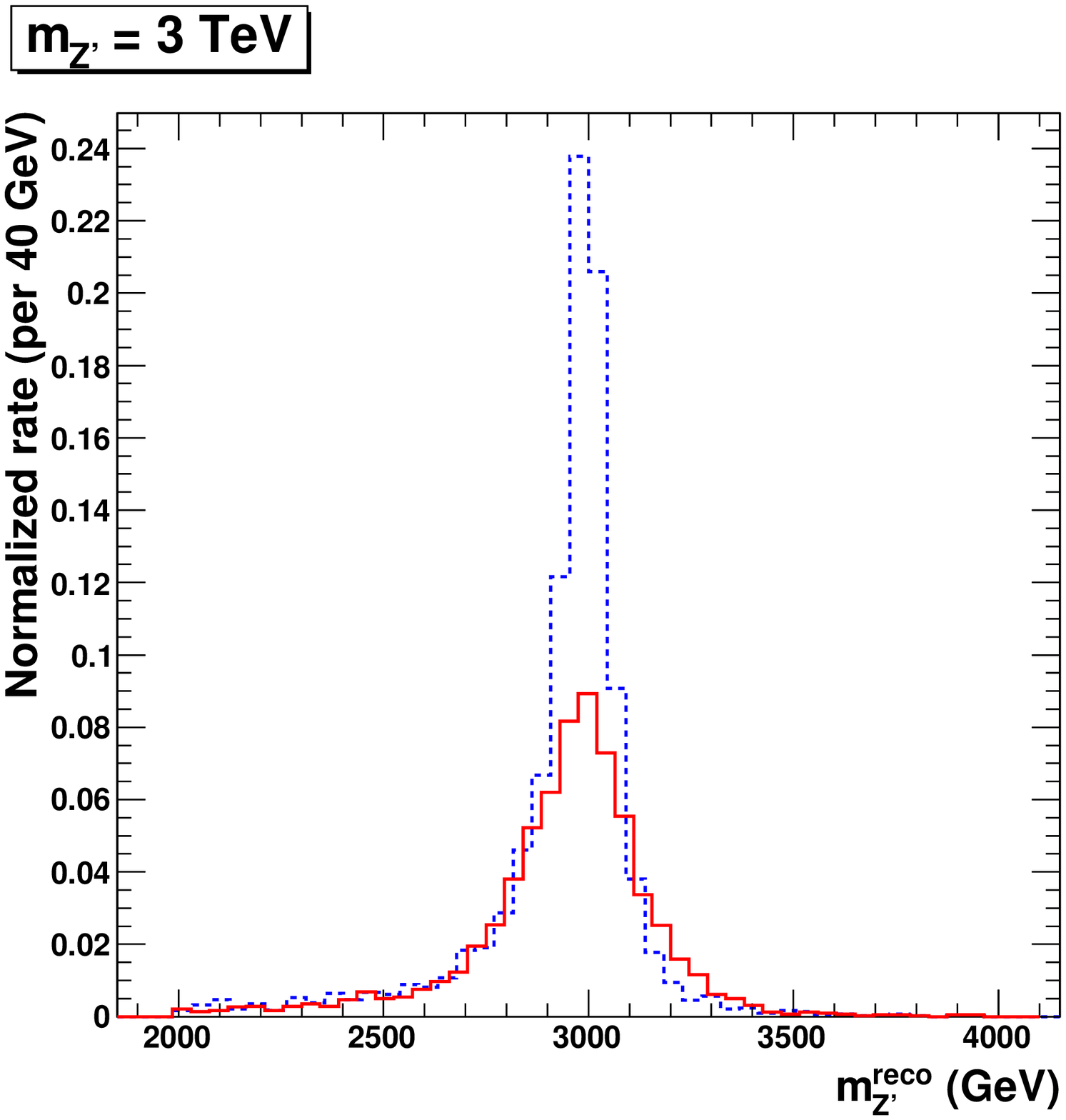}
\caption{{\it Reconstructed $Z'$ invariant mass for 2 TeV (left) and 3 TeV (right), with the Higgs mass window cut applied.  Plots are normalized to the signal rate before the Higgs mass window cut.  The solid red line uses our nominal \met\ reconstruction, and the dashed blue line uses perfect \met.}}
\label{fig:ditau4}
\end{center}
\end{figure}

The reconstructed Higgs mass is shown in Fig.~\ref{fig:ditau3}, including a comparison to a reconstruction using a perfect \met\ measurement (tracing the true sum of unsmeared neutrinos).\footnote{The plot sums over all ditau decay channels, but the individual distributions are not very different.  While this is perhaps the first time such an estimate has been made at high Higgs $p_T$, we point out for comparison the result for Higgs production from vector boson fusion in~\cite{Aad:2009wy}.  This achieves 10\% mass resolution in both dileptonic and semileptonic decay modes.}  The signal has a well-defined peak in both cases.  However, in going from a perfect \met\ measurement to a smeared measurement, the peak broadens and develops a substantive high-mass tail.  The tail in fact extends well beyond the range of the plots.  It is fed mainly by cases where the direction of the \met\ vector falls outside of the wedge defined by the two taus in the transverse plane.  In such cases, one of the neutrinos is reconstructed {\it anti}-parallel to one of the taus, and the mass broadly overshoots.\footnote{It would likely be useful to consider more advanced reconstruction methods that correct for \met\ direction mismeasurements, for example re-aligning \met\ with the closest tau in pathological cases.  We also emphasize that aside from this ``out-of-wedge'' case, we find that the mass reconstruction for nearly-aligned taus is quite stable.  This is in contrast to the other extreme where the taus are moving in nearly opposite directions, and the sum of neutrino energies can become ambiguous.}  We take as a nominal Higgs mass window $m_h^{\rm reco} = [100,300]$ GeV, which encompasses the bulk of the peak while staying somewhat higher than the $Z$ mass and recapturing some of the lost high-mass events.  Reconstruction of the $Z'$ after application of this window is shown in Fig.~\ref{fig:ditau4}.

One of the major backgrounds for this analysis is $Z$+jets, with the jets faking a ditau pair.\footnote{The diboson background $ZZ\to(l^+l^-)(\tau^+\tau^-)$ is small, and is ultimately brought down to a negligible level with our Higgs mass window.  The irreducible $Zh$ background is similarly negligible.  There will also be a background where the $Z$ is produced recoiling against a quark, which subsequently emits a $W$ boson ($W$-strahlung) that decays leptonically or into a tau.  The fraction of $Z$+jets events in which this occurs is sub-percent~\cite{Rehermann:2010vq}, before demanding that the associated jet is tau-tagged.  This background ends up subleading to $Z$+jets with mistagged ditau.}  We use this background to obtain an estimate for ditau-tag fake rates from QCD jets.  We restrict to events with a reconstructed leptonic $Z$ and where the recoiling hard parton is in the central region $|\eta| < 1$. We run separate analyses for 2(3) TeV by taking $\sqrt{\hat s} = [1600,2400]([2600,3400])$ GeV and $\hat{p}_T = [800,1200]([1300,1700])$ GeV.  The ditau tag rate is defined as the fraction of these events that pass our reconstruction.  For the equivalent analyses with the signal, we simply start with all events that pass basic event reconstruction.   We show the tag and mistag rates for the individual ditau decay modes in table~\ref{tab:ditau1}.  The samples are heavily dominated by merged ditau configurations.  We see that the signal tag rates are at the $50-70\%$ level in the semileptonic and all-hadronic channels, and the mistag rates are at the $10^{-4}-10^{-3}$ level (corresponding roughly to the square of individual hadronic tau tag rates expected at lower energies~\cite{CMStau}).  Note that the background simulations run out of statistics after application of the Higgs mass window, corresponding to final efficiencies of less than about $10^{-5}$.  The final signal efficiencies after application of the mass window are about $30-40$\%.

\begin{table}
 \centering
\begin{tabular}{|c|c|c|c|c|c|c|} \hline

      & \multicolumn{3}{|c|}{ 2 TeV } & \multicolumn{3}{|c|}{ 3 TeV } \\ \hline 
      & dilep & semilep & all-had     & dilep & semilep & all-had     \\ \hline \hline

signal ditau tag           &  86\%   & 50\%          & 49\%          & 98\%     & 51\%          & 70\%  \\
+ Higgs window             &  48\%     & 30\%          & 30\%          & 47\%     & 29\%          & 41\% \\ \hline
ditau mistag $gZ \; (qZ)$  &  - (-) & 0.02\%(0.03\%) & 0.02\%(0.07\%) & - (-) & 0.03\%(0.03\%) & $<0.01$\%(0.07\%) \\ \hline
  
\end{tabular}
\caption{{\it Efficiencies (in percent) for reconstruction of ditaus.  For the signal, the efficiencies are determined channel-by-channel.  For the background, which does not have physical taus, the efficiencies represent the percentage of all events passing into each channel.}}
\label{tab:ditau1}
\end{table}

We now turn to discovery reach.  For our signal box, we use the Higgs mass window above, and define $Z'$ mass windows of $[1650,2350]$ GeV and $[2500,3500]$ GeV.  For 2 and 3 TeV resonances, we find negligible background rate, amounting to less than 1 event even after 300~fb$^{-1}$.  (The final $Z'\to Zh$ reconstruction efficiency, after all cuts, is about $25-30\%$ times the individual $Z$ and Higgs $BR$s.)  Discovery is therefore purely determined by our requirement that at least 10 events are observed.  We combine all three ditau decay channels, and display the reach in table~\ref{tab:ditau2}.  Resonances with $\sigma \times BR(Zh)$ of a few 10's of fb can be discovered with reasonable LHC luminosities, roughly independent of mass.  These are fairly high cross sections for multi-TeV resonances.  As anticipated, the search sensitivity is limited by the low branching fraction of $Zh\to(l^+l^-)(\tau^+\tau^-)$.  However, this may still be an interesting mode to pursue for nonstandard Higgses or other scalars.  Extending the search to lower $Z'$ masses, where the cross section is expected to be larger, should also be possible.

\begin{table}
 \centering
\begin{tabular}{|c|c|c|} \hline
     & $M_{Z'} = 2 \ {\rm TeV}$ & $M_{Z'} = 3 \ {\rm TeV}$ \\ \hline \hline 
${\cal L} = 100\ {\rm fb}^{-1}$ & 77 fb & 70 fb \\
${\cal L} = 300\ {\rm fb}^{-1}$ & 26 fb& 23 fb \\ \hline 
\end{tabular}
\caption{{\it Discovery reach for $Z' \to Zh \to (l^+ l^-)(\tau^+\tau^-)$.  The numbers denote the minimum discoverable $\sigma(q \bar q\to Z')\times BR(Z' \to Zh)$ at the given integrated luminosity.  We assume $BR(h\to\tau^+\tau^-) = 0.068$, as for a standard Higgs at 120 GeV.}}
\label{tab:ditau2}
\end{table}

\subsection{Hadronic $Z$}\label{Zjjhtt}

As in the previous analysis, we start by looking for isolated leptons, and then try to find leptons mutually isolated from hadronic taus.  The event should have no good dileptonic $Z$ candidate.  The remaining particles are then clustered with $R=0.4$ C/A in search for hadronic tau-jets and ditau-jets.  Of the identified taus (or leptons), two should be on the same side of the event, and their vector sum (neglecting neutrinos) should again have $|\eta| < 1.5$.

Subsequently, we look for the hadronic $Z$ using jet substructure methods.  All particles not associated with taus are reclustered into $R=1.4$ quasi-hemispheric fat-jets.  The hardest of these should have $|\eta| < 1.5$, $p_T > 600$ GeV, and be greater than $\pi/2$ away from the taus in $\phi$.  This is declustered into two subjets as in~\cite{Butterworth:2008iy}.  The mass of the summed subjets is then required to be in the window $[75,105]$ GeV.  The final ``$Z$-tag'' rate is about 70\%, with mistag rates at the 10\% level or smaller.  For more details on this procedure and its performance (studied for the nearly identical case of $W$-jets), see~\cite{Katz:2010mr}.

\begin{figure}[t]
\begin{center} 
\epsfxsize=0.47\textwidth\epsfbox{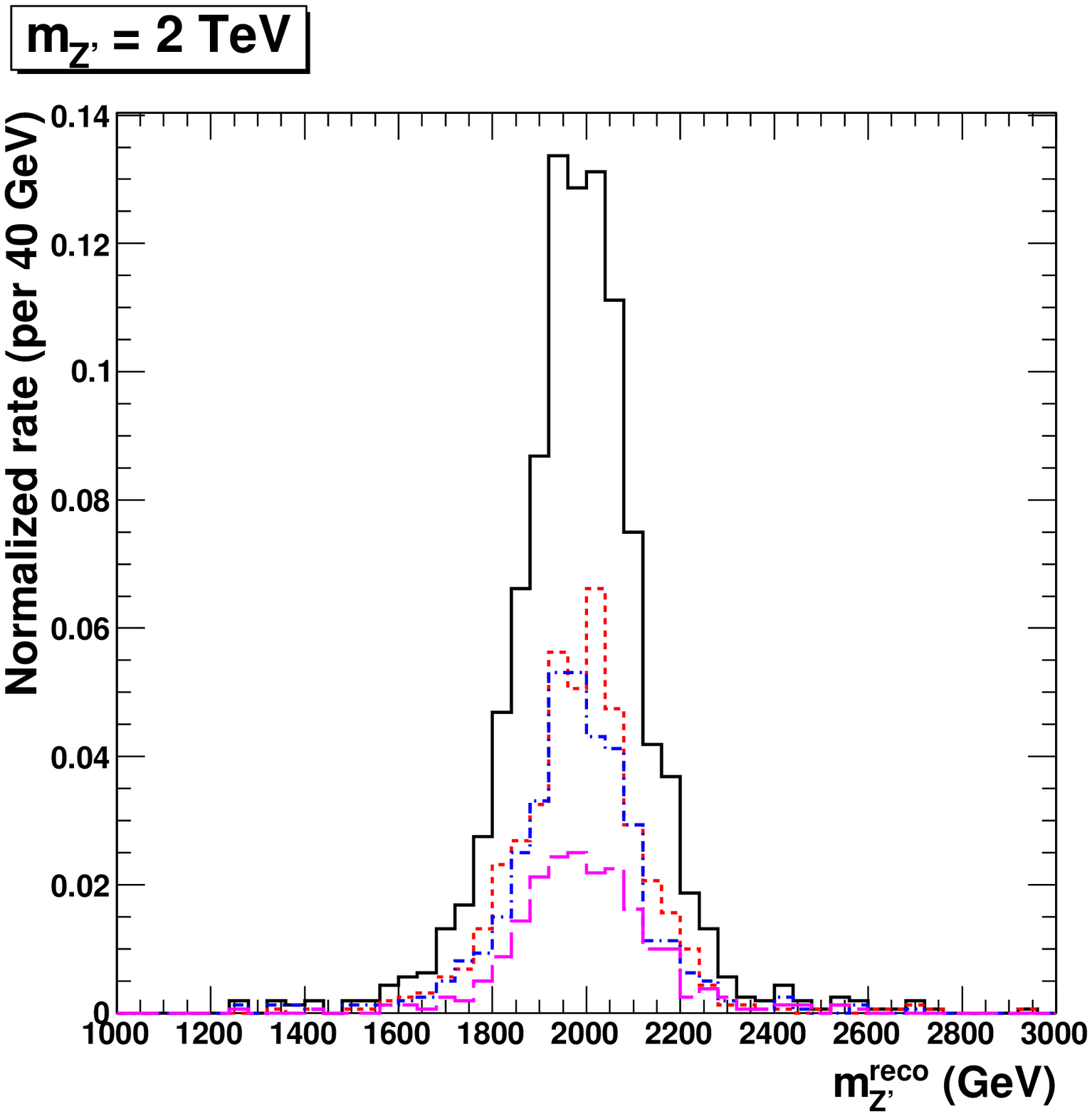}
\epsfxsize=0.47\textwidth\epsfbox{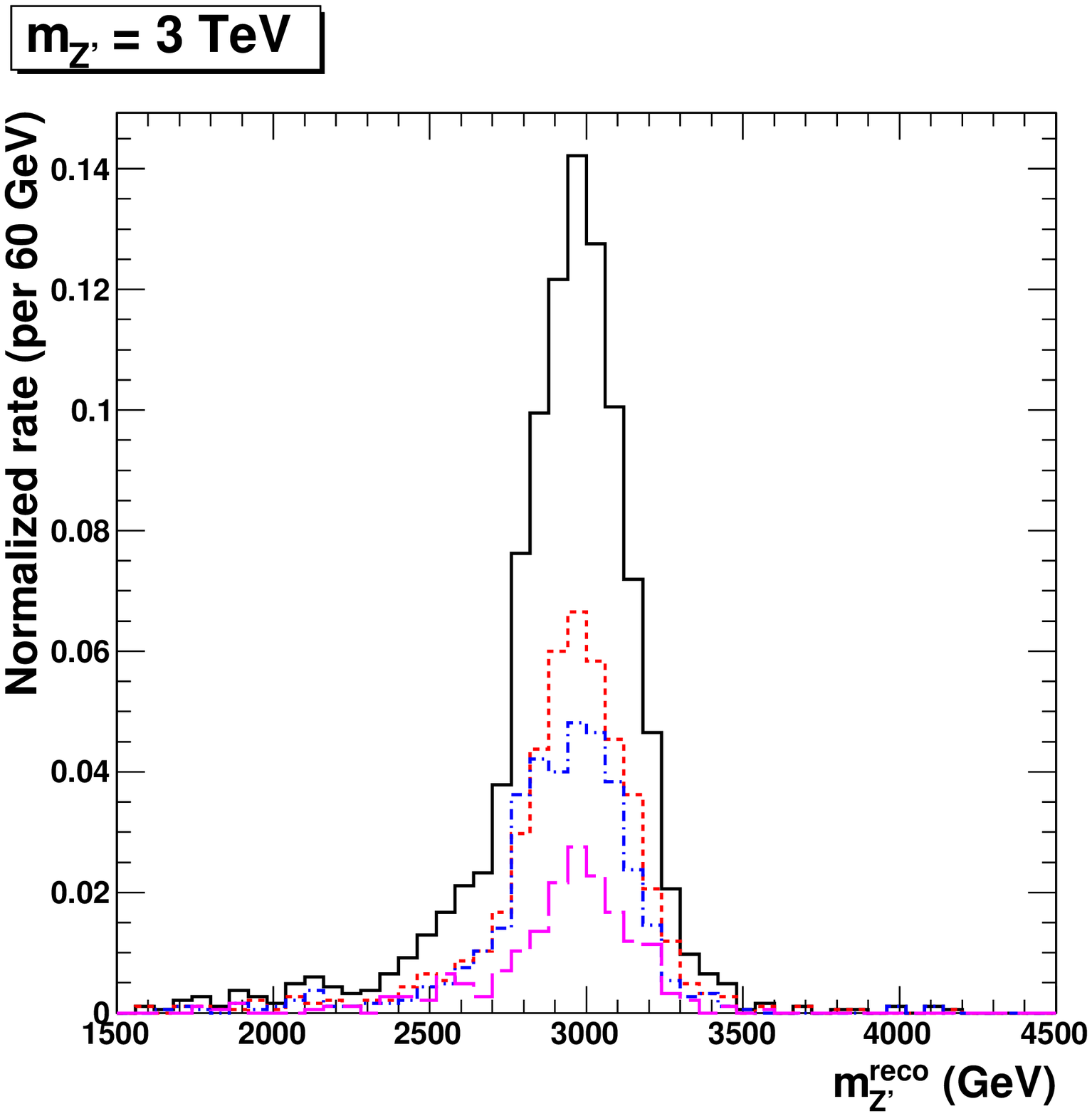}
\caption{{\it Reconstructed $Z'$ invariant mass in the hadronic $Z$ channel (black solid), and broken down into contributions from dileptonic ditau (pink long-dashed), semileptonic ditau (blue dot-dashed), and all-hadronic ditau (red short-dashed).}}
\label{fig:ditau5}
\end{center}
\end{figure}

To perform the resonance search, we use the $[100,300]$ GeV mass window for the Higgs, as above, and for the 2(3) TeV $Z'$ mass window we again use $[1650,2350]([2500,3500])$ GeV.  The reconstructed $Z'$ mass peaks are shown in Fig.~\ref{fig:ditau5}.

The backgrounds for this analysis can be formidable, especially in the all-hadronic ditau mode, where we subject ourselves to dijet-induced fakes.\footnote{Triggering on all-hadronic signal events may not be guaranteed, even with a dedicated hadronic tau trigger.  However, these events are extremely energetic, with leading jets usually well in excess of 500 GeV.  The total QCD rate above this $p_T$ is sub-Hertz at $10^{-34}$~cm$^{-1}$~s$^{-1}$ luminosity.}  However, even in this mode we find that we may achieve enough discriminating power, reducing the dijet contribution by a factor of $O(10^{-6})$ with the combination of cuts on the hadronic $Z$ and ditau Higgs.

\begin{figure}[t]
\begin{center} 
\epsfxsize=0.47\textwidth\epsfbox{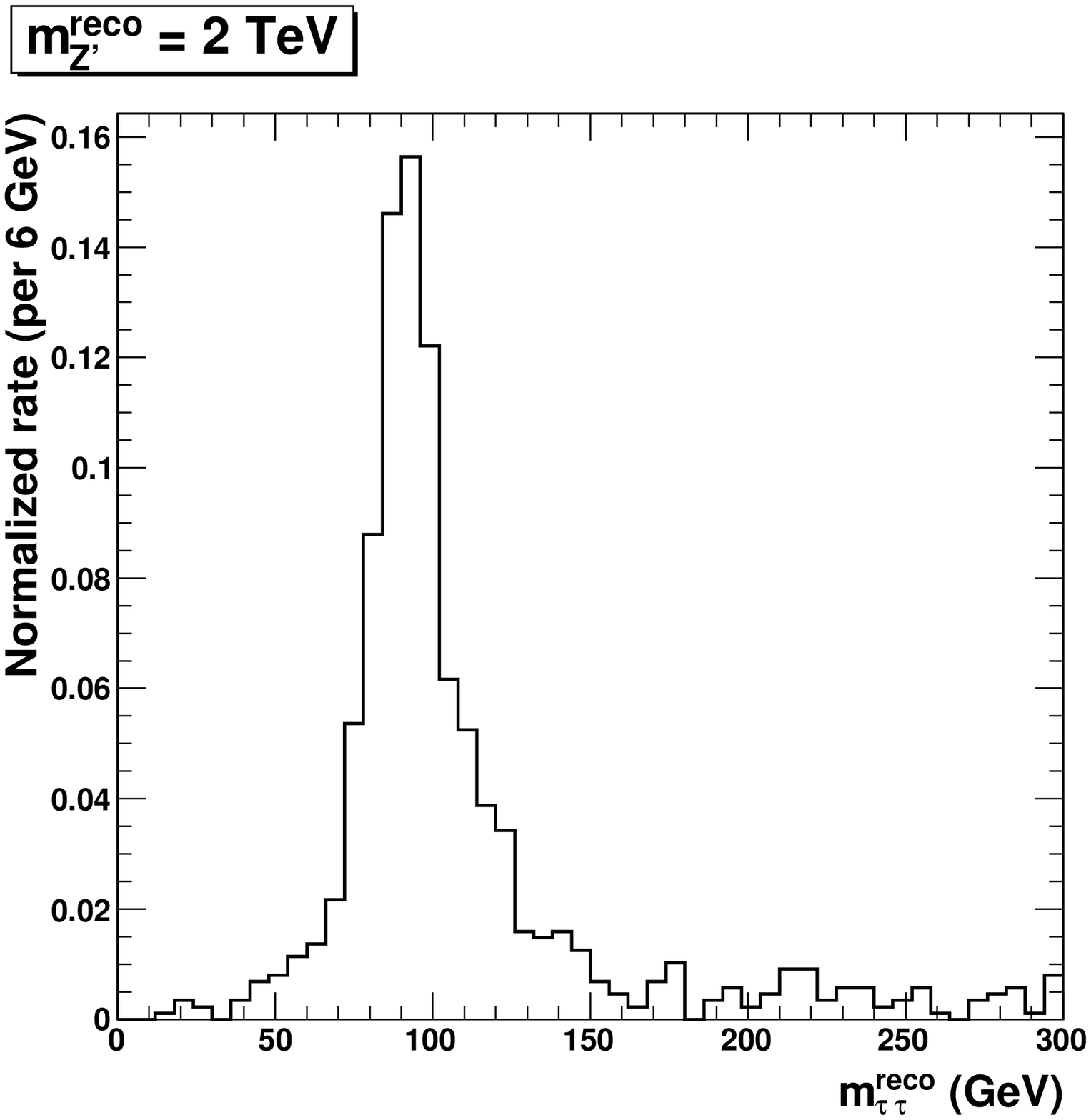}
\epsfxsize=0.47\textwidth\epsfbox{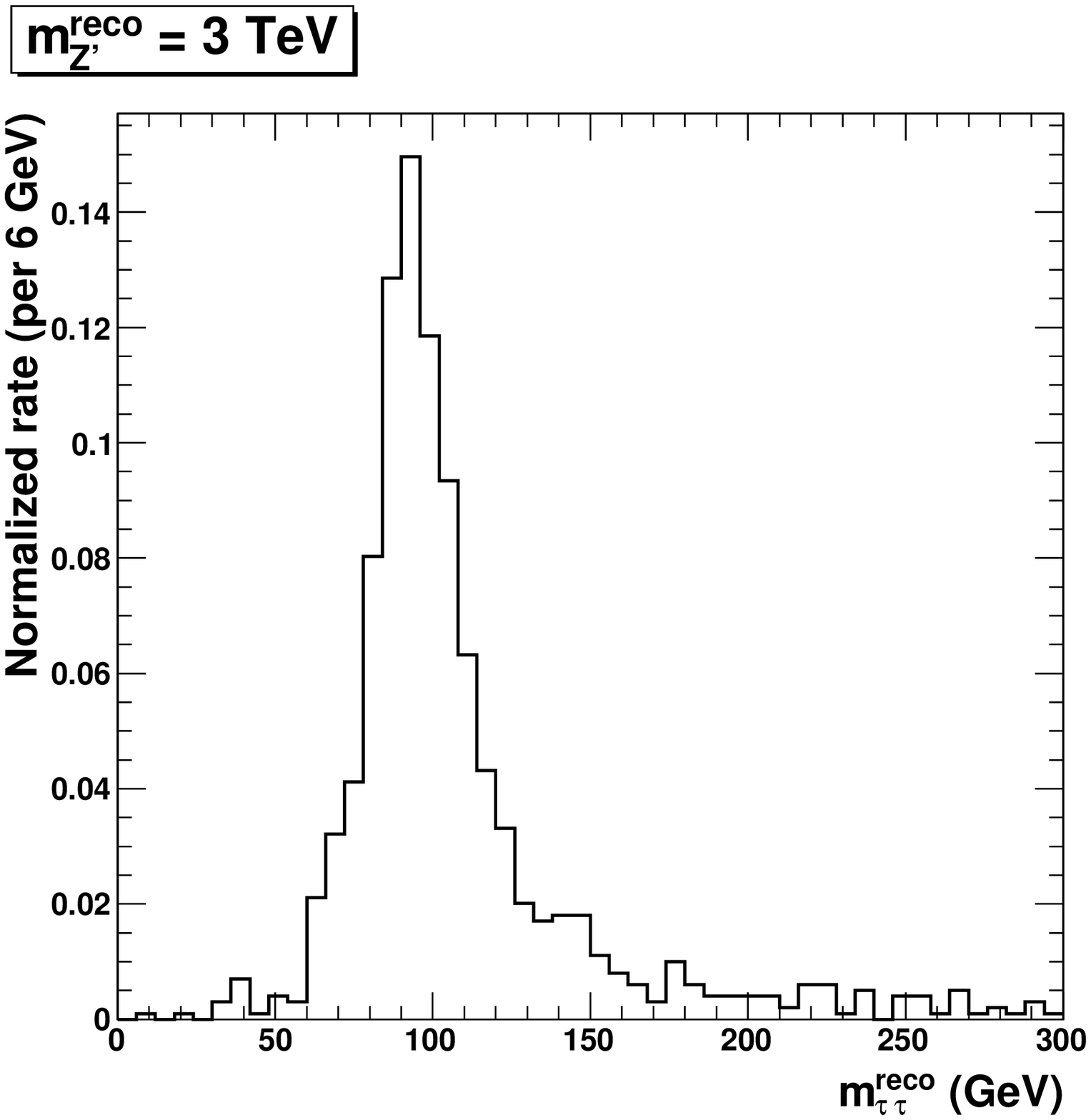}
\caption{{\it Fully reconstructed ditau invariant mass from the $(Z\to\tau^+\tau^-)$+jets background.}}
\label{fig:ditau6}
\end{center}
\end{figure}

Another important background is $Z$+jets, where this time the $Z$ decays to taus and the jets fake a hadronic $Z$.  This is the only major background with large production cross section and a genuine ditau pair.  We display the reconstructed ditau invariant mass for this background in figure~\ref{fig:ditau6}.  Our Higgs mass window cut removes about 3/4 of the events.

We also consider $t\bar t$ and $W$+jets ($W$-strahlung) backgrounds.  The former can be dangerous, for example, when one of the tops decays hadronically and the other leptonically or into a tau.  Such a $t\bar t$ event can closely mimic the signal when the $b$-jet on the hadronic side is soft (and thrown away by the hadronic $Z$-tagger), and the $b$-jet on the leptonic side is misidentified as a tau.  Final-state $W$-strahlung can occur in any process that produces a hard final state quark, in this case simple 2~$\to$~2 parton scattering.  This kind of emission greatly enhances the chance that a generic dijet event passes our analysis cuts if the emitted $W$ decays leptonically or into a tau.  However, this competes with the fact that the emission is fairly rare.

Finally, we include Standard Model diboson backgrounds, including the irreducible $Zh$.  These are all subleading.

We list the signal efficiencies and major background cross sections for the different ditau modes and $Z'$ masses in table~\ref{tab:bg}.  The backgrounds are all quite small, at a level much less than a fb, even for the all-hadronic mode.  Indeed, the dominant backgrounds in reality will likely be instrumental in nature, but determining this will require a more advanced simulation or data-driven study.\footnote{Accurate estimation of the fully QCD background is further complicated by the large statistics involved.  Our final QCD samples (consisting of several 10's of millions of events) represent event weights of about 7~ab, or 0.7 for 100~fb$^{-1}$ luminosity and 2 for 300~fb$^{-1}$.  Consequently, a quoted cross section of 7~ab is more accurately $O$(7~ab), and a quoted cross section of 0~ab is more accurately $\lsim 7$~ab.  We infer from intermediate stages of the analysis that the dileptonic and semileptonic cross sections stay well below the all-hadronic, and take ``0~ab'' to be reasonable at our level of precision.  However, for the 3 TeV all-hadronic cross section, which is represented by a single simulation event, we conservatively increase the number to 15~ab for the purpose of calculating discovery reach.}

\begin{table}
\centering

\begin{tabular}{|l|c|c|c|c|c|c|} \hline

      & \multicolumn{3}{|c|}{ 2 TeV } & \multicolumn{3}{|c|}{ 3 TeV } \\ \hline 
      & dilep & semilep & all-had     & dilep & semilep & all-had     \\ \hline \hline

Signal Eff.     &  25\% (32\%) &  12\% (16\%)  & 16\% (22\%)  &  25\% (36\%)  &  14\% (20\%)  &  19\% (29\%)  \\ \hline

$\sigma$(dijet QCD)     &  0 ab (0 ab) &  0 (39)  & 140 (1560) &  0 (0) &  0 (37)  &  7 (465)  \\
$\sigma$($Z$+jets)      &  3 (13)      &  6 (42)  &  10 (52)   &  0 (1) &  1 (2)   &  1 (2)    \\
$\sigma(t\bar t)$       &  0 (8)       &  2 (26)  &   6 (34)   &  0 (0) &  1 (3)   &  0 (4)    \\
$\sigma$($W$-strahlung) &  0 (0)       & 10 (144) &   0 (0)    &  0 (0) &  0 (84)  &  0 (0)    \\
$\sigma$(SM diboson)    &  1 (1)       &  2 (3)   &   2 (3)    &  0 (0) &  0 (0)   &  0 (0)    \\ \hline

\end{tabular}
\caption{{\it Final signal efficiencies (channel-by-channel, without incorporating branching fractions) and background cross sections, in ab, after(before) the Higgs mass cut.  Here, ``0'' indicates cross sections which we estimate to be less than 1~ab, or for which we have otherwise run out of statistics in our simulations (see text).}}
\label{tab:bg}
\end{table}

We list the discovery reach estimates in table~\ref{tab:ditau3}, broken down into cases without/with the inclusion of the all-hadronic mode.  While all-hadronic is likely the most experimentally challenging, we have seen that its backgrounds appear to be controllable, and that including it roughly doubles the statistics.  However, in the event that instrumental backgrounds make this mode non-viable, the combined dileptonic and semileptonic analyses can still give good, even superior sensitivity, especially at very high integrated luminosity.

\begin{table}
 \centering
\begin{tabular}{|c|c|c|c|c|} \hline
     & $M_{Z'} = 2 \ {\rm TeV}$ & $M_{Z'} = 3 \ {\rm TeV}$ & $M_{Z'} = 2 \ {\rm TeV}$ ($\ge 1$ lepton) & 
$M_{Z'} = 3 \ {\rm TeV}$ ($\ge 1$ lepton) \\ \hline \hline 
${\cal L} = 30\ {\rm fb}^{-1} $  & 66 fb & 35 fb & 76 fb & 63 fb \\
${\cal L} = 100\ {\rm fb}^{-1} $ & 32 fb & 12 fb & 26 fb & 21 fb \\
${\cal L} = 300\ {\rm fb}^{-1}$  & 17 fb &  6 fb & 13 fb &  7 fb \\ \hline
S/B = 1                         & 23 fb &  4 fb &  6 fb &  1 fb \\ \hline 
\end{tabular}
\caption{{\it Discovery reach for $Z' \to Zh \to (q\bar q)(\tau^+\tau^-)$.  
The numbers denote the minimum discoverable $\sigma(q \bar q\to Z')\times BR(Z' \to Zh)$ at the given integrated luminosity, or 
that which is required to match the background cross section after all cuts.  
We assume $BR(h\to\tau^+\tau^-) = 0.068$, as for a standard Higgs at 120 GeV.  The last two columns use only the channels
with at least one leptonic tau, omitting the all-hadronic channel.}}
\label{tab:ditau3}
\end{table}

As expected, the overall gain in sensitivity is significant relative to the $(l^+l^-)(\tau^+\tau^-)$ channel.  For example, discovery of a 3 TeV $Z'$ with $\sigma\times BR(Zh)$ of about 6-7~fb is possible at 300~fb$^{-1}$ luminosity (versus 23~fb with the leptonic $Z$).   These can be compared to the results in~\cite{Katz:2010mr}, utilizing $h\to b\bar b$ in association with a leptonic or invisible $Z$, which respectively could probe down to about 2.5~fb and 1.5~fb.  For 2 TeV resonances, the difference in sensitivities is more modest, with $(q\bar q)(\tau^+\tau^-)$ (omitting all-hadronic ditau) only about 50\% less sensitive than the non-$b$-tagged $(l^+l^-)(b\bar b)$.  In particular, the $Z'$ states of the warped 5D models of~\cite{Agashe:2003zs,Agashe:2007ki} may be simultaneously visible at 2 TeV, with production cross section of about 15~fb, in the three $Zh$ modes $(l^+l^-)(b\bar b)$, $(\nu\bar\nu)(b\bar b)$, and $(q\bar q)(\tau^+\tau^-)$.

Finally, we note that all of our results in this subsection can also be trivially applied to $W'\to Wh\to(q\bar q')(\tau^+\tau^-)$, which has almost identical structure.

\section{Conclusions}
\label{sec:outlook}


When a light Higgs is generated in a multi-TeV process, it can subsequently decay into a pair of taus and manifest itself in the detector as a ditau-jet.  While this decay channel is expected to be subdominant, it is also extremely distinctive.  Both the Higgs and the tau are color-singlets, so any hadrons produced in the tau decays are highly beamed into one or two very small regions of the detector without any surrounding radiation produced in the final state.  This unusual configuration should in principle become progressively easier to spot at higher energies.  In this paper, we have explored the possibility of capitalizing on this fact in order to construct a powerful high-$p_T$ Higgs tag, and to determine if $h\to\tau^+\tau^-$ can have any role to play in the search for heavy new physics at the LHC.

In order to identify tau pairs with $\Delta R_{\tau\tau} \lsim 0.4$ and one or two hadronic decays, we extend the usual tau algorithms to incorporate ``mutual isolation,'' where the inner activity cone of one tau is deleted from the isolation annulus of its partner.  We have found that for ditaus at $p_T \sim$~TeV we can achieve tag and mistag rates that are roughly the square of what has been estimated by the LHC experimentalists for single taus at $p_T \sim 100$~GeV.  Specifically, we estimate that $O$(50\%) tag rate and $O(10^{-4}-10^{-3})$ mistag rates should be possible.  Of course, in this study we were not able to faithfully model instrumental effects, which are likely very important.  A more detailed detector study is warranted to verify the physics picture which we have laid out here.

We have also demonstrated that the ditau pair mass can be reconstructed, with the Higgs appearing as a broad but well-localized bump.  This is perhaps counterintuitive, as it is known that Higgs mass reconstruction becomes badly ambiguous in the {\it anti}-collinear limit, due to a degeneracy in the apportioning of $\vec{\displaystyle{\not}E}_T$ between the two taus.  However, taking the collinear limit and assuming that all of the neutrinos are moving in roughly the same direction as the visible tau products, this ambiguity is no longer so severe since we know the approximate sum of neutrino energies.  Nevertheless, fluctuations in visible energy measurements can cause the missing energy vector to point outside of the 2D wedge in the transverse plane defined by the visible taus, naively leading to reconstructed neutrinos oriented toward the wrong side of the event.  In this case, the reconstructed ditau mass can be highly overestimated.  While this population of misreconstructed events does not cause major problems for us, it seems likely that even a slightly modified method could perform much better.

We applied the ditau-jet tag and invariant mass reconstruction to the simple new physics process of a multi-TeV $Z'$ decaying to $Zh$.  (Some of our results also trivially extend to $W'\to Wh$.)  The low-statistics channel $(l^+l^-)(\tau^+\tau^-)$ mainly served to calibrate our ditau techniques, but is largely background-free and has sensitivity to resonances with $\sigma\times BR(Zh)$ of a few 10's of fb after running the LHC for several hundred fb$^{-1}$.  The channel $(q\bar q)(\tau^+\tau^-)$ is more promising from the perspective of rate, but faces enormous backgrounds at the outset.  Nonetheless, we have shown that the combination of ditau-tagging, Higgs mass reconstruction, and hadronic $Z$-tagging (as in~\cite{Katz:2010mr}) can potentially reduce the background to a manageable level while preserving most of the signal.  We estimate that this channel has sensitivity at the 10~fb level or lower, competing with the more conventional channels $(l^+l^-)(b\bar b)$ and $(\nu\bar\nu)(b\bar b)$~\cite{Katz:2010mr}, especially around 2 TeV.

The techniques which we discuss here can also be applied in a broader new physics context.  In addition to resonant production of the Higgs, we might consider Higgses produced in SUSY cascades or in the decays of other pair-produced heavy particles.  New scalars besides the Higgs are another possible target.  For example, a light scalar ($m \ll 100$ GeV) decaying into tau pairs could be identified as a ditau-jet within events of much lower energies.

\acknowledgments{We are grateful to Kaustubh Agashe, Sarah Demers, Petar Maksimovic, Frank Paige, and Raman Sundrum for useful discussions.  The authors are also grateful to the organizers of the workshop ``BOOST2010'' at Oxford.  AK was partially supported by NSF grant PHY-0801323.  MS was supported in part by the DoE under grant No.\ DE-FG-02-92ER40704.  BT was supported by JHU grant No.\ 80020033 and by DoE grant No.\ DE-FG-02-91ER40676.}


\bibliography{lit}

\begin{thebibliography}{38}
\expandafter\ifx\csname natexlab\endcsname\relax\def\natexlab#1{#1}\fi
\expandafter\ifx\csname bibnamefont\endcsname\relax
  \def\bibnamefont#1{#1}\fi
\expandafter\ifx\csname bibfnamefont\endcsname\relax
  \def\bibfnamefont#1{#1}\fi
\expandafter\ifx\csname citenamefont\endcsname\relax
  \def\citenamefont#1{#1}\fi
\expandafter\ifx\csname url\endcsname\relax
  \def\url#1{\texttt{#1}}\fi
\expandafter\ifx\csname urlprefix\endcsname\relax\def\urlprefix{URL }\fi
\providecommand{\bibinfo}[2]{#2}
\providecommand{\eprint}[2][]{\url{#2}}

\bibitem[{\citenamefont{Seymour}(1994)}]{Seymour:1993mx}
\bibinfo{author}{\bibfnamefont{M.~H.} \bibnamefont{Seymour}},
  \emph{\bibinfo{title}{{Searches for New Particles Using Cone and Cluster Jet
  Algorithms: A Comparative Study}}}, \bibinfo{journal}{Z. Phys.}
  \textbf{\bibinfo{volume}{C62}}, \bibinfo{pages}{127} (\bibinfo{year}{1994}).

\bibitem[{\citenamefont{Butterworth et~al.}(2002)\citenamefont{Butterworth,
  Cox, and Forshaw}}]{Butterworth:2002tt}
\bibinfo{author}{\bibfnamefont{J.~M.} \bibnamefont{Butterworth}},
  \bibinfo{author}{\bibfnamefont{B.~E.} \bibnamefont{Cox}}, \bibnamefont{and}
  \bibinfo{author}{\bibfnamefont{J.~R.} \bibnamefont{Forshaw}},
  \emph{\bibinfo{title}{{$WW$ Scattering at the CERN LHC}}},
  \bibinfo{journal}{Phys. Rev.} \textbf{\bibinfo{volume}{D65}},
  \bibinfo{pages}{096014} (\bibinfo{year}{2002}), \eprint{hep-ph/0201098}.

\bibitem[{\citenamefont{Butterworth et~al.}(2008)\citenamefont{Butterworth,
  Davison, Rubin, and Salam}}]{Butterworth:2008iy}
\bibinfo{author}{\bibfnamefont{J.~M.} \bibnamefont{Butterworth}},
  \bibinfo{author}{\bibfnamefont{A.~R.} \bibnamefont{Davison}},
  \bibinfo{author}{\bibfnamefont{M.}~\bibnamefont{Rubin}}, \bibnamefont{and}
  \bibinfo{author}{\bibfnamefont{G.~P.} \bibnamefont{Salam}},
  \emph{\bibinfo{title}{{Jet Substructure as a New Higgs Search Channel at the
  LHC}}}, \bibinfo{journal}{Phys.Rev.Lett.} \textbf{\bibinfo{volume}{100}},
  \bibinfo{pages}{242001} (\bibinfo{year}{2008}), \eprint{0802.2470}.

\bibitem[{\citenamefont{Brooijmans}(2008)}]{Brooijmans:2008zz}
\bibinfo{author}{\bibfnamefont{G.}~\bibnamefont{Brooijmans}},
  \emph{\bibinfo{title}{{High $p_T$ Hadronic Top Quark Identification. Part I:
  Jet Mass and Ysplitter}}} (\bibinfo{year}{2008}),
  \eprint{ATL-PHYS-CONF-2008-008}.

\bibitem[{\citenamefont{Thaler and Wang}(2008)}]{Thaler:2008ju}
\bibinfo{author}{\bibfnamefont{J.}~\bibnamefont{Thaler}} \bibnamefont{and}
  \bibinfo{author}{\bibfnamefont{L.-T.} \bibnamefont{Wang}},
  \emph{\bibinfo{title}{{Strategies to Identify Boosted Tops}}},
  \bibinfo{journal}{JHEP} \textbf{\bibinfo{volume}{07}}, \bibinfo{pages}{092}
  (\bibinfo{year}{2008}), \eprint{0806.0023}.

\bibitem[{\citenamefont{Kaplan et~al.}(2008)\citenamefont{Kaplan, Rehermann,
  Schwartz, and Tweedie}}]{Kaplan:2008ie}
\bibinfo{author}{\bibfnamefont{D.~E.} \bibnamefont{Kaplan}},
  \bibinfo{author}{\bibfnamefont{K.}~\bibnamefont{Rehermann}},
  \bibinfo{author}{\bibfnamefont{M.~D.} \bibnamefont{Schwartz}},
  \bibnamefont{and} \bibinfo{author}{\bibfnamefont{B.}~\bibnamefont{Tweedie}},
  \emph{\bibinfo{title}{{Top Tagging: A Method for Identifying Boosted
  Hadronically Decaying Top Quarks}}}, \bibinfo{journal}{Phys. Rev. Lett.}
  \textbf{\bibinfo{volume}{101}}, \bibinfo{pages}{142001}
  (\bibinfo{year}{2008}), \eprint{0806.0848}.

\bibitem[{\citenamefont{Plehn et~al.}(2010{\natexlab{a}})\citenamefont{Plehn,
  Salam, and Spannowsky}}]{Plehn:2009rk}
\bibinfo{author}{\bibfnamefont{T.}~\bibnamefont{Plehn}},
  \bibinfo{author}{\bibfnamefont{G.~P.} \bibnamefont{Salam}}, \bibnamefont{and}
  \bibinfo{author}{\bibfnamefont{M.}~\bibnamefont{Spannowsky}},
  \emph{\bibinfo{title}{{Fat Jets for a Light Higgs}}}, \bibinfo{journal}{Phys.
  Rev. Lett.} \textbf{\bibinfo{volume}{104}}, \bibinfo{pages}{111801}
  (\bibinfo{year}{2010}{\natexlab{a}}), \eprint{0910.5472}.

\bibitem[{\citenamefont{Plehn et~al.}(2010{\natexlab{b}})\citenamefont{Plehn,
  Spannowsky, Takeuchi, and Zerwas}}]{Plehn:2010st}
\bibinfo{author}{\bibfnamefont{T.}~\bibnamefont{Plehn}},
  \bibinfo{author}{\bibfnamefont{M.}~\bibnamefont{Spannowsky}},
  \bibinfo{author}{\bibfnamefont{M.}~\bibnamefont{Takeuchi}}, \bibnamefont{and}
  \bibinfo{author}{\bibfnamefont{D.}~\bibnamefont{Zerwas}},
  \emph{\bibinfo{title}{{Stop Reconstruction with Tagged Tops}}}
  (\bibinfo{year}{2010}{\natexlab{b}}), \eprint{1006.2833}.

\bibitem[{\citenamefont{Rehermann and Tweedie}(2010)}]{Rehermann:2010vq}
\bibinfo{author}{\bibfnamefont{K.}~\bibnamefont{Rehermann}} \bibnamefont{and}
  \bibinfo{author}{\bibfnamefont{B.}~\bibnamefont{Tweedie}},
  \emph{\bibinfo{title}{{Efficient Identification of Boosted Semileptonic Top
  Quarks at the LHC}}} (\bibinfo{year}{2010}), \eprint{1007.2221}.

\bibitem[{\citenamefont{Butterworth et~al.}(2009)\citenamefont{Butterworth,
  Ellis, Raklev, and Salam}}]{Butterworth:2009qa}
\bibinfo{author}{\bibfnamefont{J.~M.} \bibnamefont{Butterworth}},
  \bibinfo{author}{\bibfnamefont{J.~R.} \bibnamefont{Ellis}},
  \bibinfo{author}{\bibfnamefont{A.~R.} \bibnamefont{Raklev}},
  \bibnamefont{and} \bibinfo{author}{\bibfnamefont{G.~P.} \bibnamefont{Salam}},
  \emph{\bibinfo{title}{{Discovering Baryon-Number Violating Neutralino Decays
  at the LHC}}}, \bibinfo{journal}{Phys. Rev. Lett.}
  \textbf{\bibinfo{volume}{103}}, \bibinfo{pages}{241803}
  (\bibinfo{year}{2009}), \eprint{0906.0728}.

\bibitem[{\citenamefont{Kribs et~al.}(2010)\citenamefont{Kribs, Martin, Roy,
  and Spannowsky}}]{Kribs:2009yh}
\bibinfo{author}{\bibfnamefont{G.~D.} \bibnamefont{Kribs}},
  \bibinfo{author}{\bibfnamefont{A.}~\bibnamefont{Martin}},
  \bibinfo{author}{\bibfnamefont{T.~S.} \bibnamefont{Roy}}, \bibnamefont{and}
  \bibinfo{author}{\bibfnamefont{M.}~\bibnamefont{Spannowsky}},
  \emph{\bibinfo{title}{{Discovering the Higgs Boson in New Physics Events
  using Jet Substructure}}}, \bibinfo{journal}{Phys. Rev.}
  \textbf{\bibinfo{volume}{D81}}, \bibinfo{pages}{111501}
  (\bibinfo{year}{2010}), \eprint{0912.4731}.

\bibitem[{\citenamefont{Almeida et~al.}(2009)}]{Almeida:2008yp}
\bibinfo{author}{\bibfnamefont{L.~G.} \bibnamefont{Almeida}}
  \bibnamefont{et~al.}, \emph{\bibinfo{title}{{Substructure of High-$p_T$ Jets
  at the LHC}}}, \bibinfo{journal}{Phys. Rev.} \textbf{\bibinfo{volume}{D79}},
  \bibinfo{pages}{074017} (\bibinfo{year}{2009}), \eprint{0807.0234}.

\bibitem[{\citenamefont{Almeida et~al.}(2010)\citenamefont{Almeida, Lee, Perez,
  Sterman, and Sung}}]{Almeida:2010pa}
\bibinfo{author}{\bibfnamefont{L.~G.} \bibnamefont{Almeida}},
  \bibinfo{author}{\bibfnamefont{S.~J.} \bibnamefont{Lee}},
  \bibinfo{author}{\bibfnamefont{G.}~\bibnamefont{Perez}},
  \bibinfo{author}{\bibfnamefont{G.}~\bibnamefont{Sterman}}, \bibnamefont{and}
  \bibinfo{author}{\bibfnamefont{I.}~\bibnamefont{Sung}},
  \emph{\bibinfo{title}{{Template Overlap Method for Massive Jets}}},
  \bibinfo{journal}{Phys. Rev.} \textbf{\bibinfo{volume}{D82}},
  \bibinfo{pages}{054034} (\bibinfo{year}{2010}), \eprint{1006.2035}.

\bibitem[{\citenamefont{Ellis et~al.}(2009)\citenamefont{Ellis, Vermilion, and
  Walsh}}]{Ellis:2009su}
\bibinfo{author}{\bibfnamefont{S.~D.} \bibnamefont{Ellis}},
  \bibinfo{author}{\bibfnamefont{C.~K.} \bibnamefont{Vermilion}},
  \bibnamefont{and} \bibinfo{author}{\bibfnamefont{J.~R.} \bibnamefont{Walsh}},
  \emph{\bibinfo{title}{{Techniques for Improved Heavy Particle Searches with
  Jet Substructure}}}, \bibinfo{journal}{Phys. Rev.}
  \textbf{\bibinfo{volume}{D80}}, \bibinfo{pages}{051501}
  (\bibinfo{year}{2009}), \eprint{0903.5081}.

\bibitem[{\citenamefont{Krohn et~al.}(2010)\citenamefont{Krohn, Thaler, and
  Wang}}]{Krohn:2009th}
\bibinfo{author}{\bibfnamefont{D.}~\bibnamefont{Krohn}},
  \bibinfo{author}{\bibfnamefont{J.}~\bibnamefont{Thaler}}, \bibnamefont{and}
  \bibinfo{author}{\bibfnamefont{L.-T.} \bibnamefont{Wang}},
  \emph{\bibinfo{title}{{Jet Trimming}}}, \bibinfo{journal}{JHEP}
  \textbf{\bibinfo{volume}{02}}, \bibinfo{pages}{084} (\bibinfo{year}{2010}),
  \eprint{0912.1342}.

\bibitem[{\citenamefont{Thaler and Van~Tilburg}(2010)}]{Thaler:2010tr}
\bibinfo{author}{\bibfnamefont{J.}~\bibnamefont{Thaler}} \bibnamefont{and}
  \bibinfo{author}{\bibfnamefont{K.}~\bibnamefont{Van~Tilburg}},
  \emph{\bibinfo{title}{{Identifying Boosted Objects with N-subjettiness}}}
  (\bibinfo{year}{2010}), \eprint{1011.2268}.

\bibitem[{\citenamefont{Katz et~al.}(2010)\citenamefont{Katz, Son, and
  Tweedie}}]{Katz:2010mr}
\bibinfo{author}{\bibfnamefont{A.}~\bibnamefont{Katz}},
  \bibinfo{author}{\bibfnamefont{M.}~\bibnamefont{Son}}, \bibnamefont{and}
  \bibinfo{author}{\bibfnamefont{B.}~\bibnamefont{Tweedie}},
  \emph{\bibinfo{title}{{Jet Substructure and the Search for Neutral Spin-One
  Resonances in Electroweak Boson Channels}}} (\bibinfo{year}{2010}),
  \eprint{1010.5253}.

\bibitem[{\citenamefont{Agashe et~al.}(2003)\citenamefont{Agashe, Delgado, May,
  and Sundrum}}]{Agashe:2003zs}
\bibinfo{author}{\bibfnamefont{K.}~\bibnamefont{Agashe}},
  \bibinfo{author}{\bibfnamefont{A.}~\bibnamefont{Delgado}},
  \bibinfo{author}{\bibfnamefont{M.~J.} \bibnamefont{May}}, \bibnamefont{and}
  \bibinfo{author}{\bibfnamefont{R.}~\bibnamefont{Sundrum}},
  \emph{\bibinfo{title}{{RS1, Custodial Isospin and Precision Tests}}},
  \bibinfo{journal}{JHEP} \textbf{\bibinfo{volume}{08}}, \bibinfo{pages}{050}
  (\bibinfo{year}{2003}), \eprint{hep-ph/0308036}.

\bibitem[{\citenamefont{Arkani-Hamed et~al.}(2001)\citenamefont{Arkani-Hamed,
  Cohen, and Georgi}}]{ArkaniHamed:2001nc}
\bibinfo{author}{\bibfnamefont{N.}~\bibnamefont{Arkani-Hamed}},
  \bibinfo{author}{\bibfnamefont{A.~G.} \bibnamefont{Cohen}}, \bibnamefont{and}
  \bibinfo{author}{\bibfnamefont{H.}~\bibnamefont{Georgi}},
  \emph{\bibinfo{title}{{Electroweak Symmetry Breaking from Dimensional
  Deconstruction}}}, \bibinfo{journal}{Phys. Lett.}
  \textbf{\bibinfo{volume}{B513}}, \bibinfo{pages}{232} (\bibinfo{year}{2001}),
  \eprint{hep-ph/0105239}.

\bibitem[{\citenamefont{Schmaltz and Tucker-Smith}(2005)}]{Schmaltz:2005ky}
\bibinfo{author}{\bibfnamefont{M.}~\bibnamefont{Schmaltz}} \bibnamefont{and}
  \bibinfo{author}{\bibfnamefont{D.}~\bibnamefont{Tucker-Smith}},
  \emph{\bibinfo{title}{{Little Higgs Review}}}, \bibinfo{journal}{Ann. Rev.
  Nucl. Part. Sci.} \textbf{\bibinfo{volume}{55}}, \bibinfo{pages}{229}
  (\bibinfo{year}{2005}), \eprint{hep-ph/0502182}.

\bibitem[{\citenamefont{Langacker}(2009)}]{Langacker:2008yv}
\bibinfo{author}{\bibfnamefont{P.}~\bibnamefont{Langacker}},
  \emph{\bibinfo{title}{{The Physics of Heavy $Z^\prime$ Gauge Bosons}}},
  \bibinfo{journal}{Rev. Mod. Phys.} \textbf{\bibinfo{volume}{81}},
  \bibinfo{pages}{1199} (\bibinfo{year}{2009}), \eprint{0801.1345}.

\bibitem[{\citenamefont{Garcia et~al.}(2004)\citenamefont{Garcia, Lechowski,
  Ros, and Rousseau}}]{Atl:LH}
\bibinfo{author}{\bibfnamefont{J.~E.} \bibnamefont{Garcia}},
  \bibinfo{author}{\bibfnamefont{M.}~\bibnamefont{Lechowski}},
  \bibinfo{author}{\bibfnamefont{E.}~\bibnamefont{Ros}}, \bibnamefont{and}
  \bibinfo{author}{\bibfnamefont{D.}~\bibnamefont{Rousseau}}
  (\bibinfo{collaboration}{The Atlas}), \emph{\bibinfo{title}{{Search for the
  Decays $Z_H \to Zh$ and $W_H \to Wh$ in the Little Higgs Model Assuming $m(h)
  = 120$ GeV}}} (\bibinfo{year}{2004}), \eprint{ATL-PHYS-2004-001}.

\bibitem[{\citenamefont{Agashe et~al.}(2007)}]{Agashe:2007ki}
\bibinfo{author}{\bibfnamefont{K.}~\bibnamefont{Agashe}} \bibnamefont{et~al.},
  \emph{\bibinfo{title}{{LHC Signals for Warped Electroweak Neutral Gauge
  Bosons}}}, \bibinfo{journal}{Phys. Rev.} \textbf{\bibinfo{volume}{D76}},
  \bibinfo{pages}{115015} (\bibinfo{year}{2007}), \eprint{0709.0007}.

\bibitem[{\citenamefont{Gallicchio and Schwartz}(2010)}]{Gallicchio:2010sw}
\bibinfo{author}{\bibfnamefont{J.}~\bibnamefont{Gallicchio}} \bibnamefont{and}
  \bibinfo{author}{\bibfnamefont{M.~D.} \bibnamefont{Schwartz}},
  \emph{\bibinfo{title}{{Seeing in Color: Jet Superstructure}}},
  \bibinfo{journal}{Phys. Rev. Lett.} \textbf{\bibinfo{volume}{105}},
  \bibinfo{pages}{022001} (\bibinfo{year}{2010}), \eprint{1001.5027}.

\bibitem[{\citenamefont{Falkowski et~al.}(2010)\citenamefont{Falkowski, Krohn,
  Wang, Shelton, and Thalapillil}}]{Falkowski:2010hi}
\bibinfo{author}{\bibfnamefont{A.}~\bibnamefont{Falkowski}},
  \bibinfo{author}{\bibfnamefont{D.}~\bibnamefont{Krohn}},
  \bibinfo{author}{\bibfnamefont{L.-T.} \bibnamefont{Wang}},
  \bibinfo{author}{\bibfnamefont{J.}~\bibnamefont{Shelton}}, \bibnamefont{and}
  \bibinfo{author}{\bibfnamefont{A.}~\bibnamefont{Thalapillil}},
  \emph{\bibinfo{title}{{Unburied Higgs}}} (\bibinfo{year}{2010}),
  \eprint{1006.1650}.

\bibitem[{\citenamefont{Heldmann and Covalli}(2006)}]{TauCalAtlas}
\bibinfo{author}{\bibfnamefont{M.}~\bibnamefont{Heldmann}} \bibnamefont{and}
  \bibinfo{author}{\bibfnamefont{D.}~\bibnamefont{Covalli}}
  (\bibinfo{collaboration}{The Atlas}), \emph{\bibinfo{title}{An improved
  tau-identification for the atlas experiment}} (\bibinfo{year}{2006}),
  \eprint{ATL-PHYS-PUB-2006-008}.

\bibitem[{\citenamefont{Richter-Was and Szymocha}(2005)}]{TauTrackAtlas}
\bibinfo{author}{\bibfnamefont{E.}~\bibnamefont{Richter-Was}} \bibnamefont{and}
  \bibinfo{author}{\bibfnamefont{T.}~\bibnamefont{Szymocha}}
  (\bibinfo{collaboration}{The Atlas}), \emph{\bibinfo{title}{{Hadronic Tau
  Identification with Track Based Approach: The $Z \to \tau \tau$, $W \to \tau
  \nu $ and Dijet Events from DC1 Data Samples}}} (\bibinfo{year}{2005}),
  \eprint{ATL-PHYS-PUB-2005-005}.

\bibitem[{\citenamefont{{The CMS Collaboration}}(2009)}]{CMStau}
\bibinfo{author}{\bibnamefont{{The CMS Collaboration}}},
  \emph{\bibinfo{title}{{CMS Strategies for Tau Reconstruction and
  Identification Using Particle Flow Techniques}}} (\bibinfo{year}{2009}),
  \eprint{CMS PAS PFT-08-001}.

\bibitem[{\citenamefont{Dokshitzer et~al.}(1997)\citenamefont{Dokshitzer,
  Leder, Moretti, and Webber}}]{Dokshitzer:1997in}
\bibinfo{author}{\bibfnamefont{Y.~L.} \bibnamefont{Dokshitzer}},
  \bibinfo{author}{\bibfnamefont{G.~D.} \bibnamefont{Leder}},
  \bibinfo{author}{\bibfnamefont{S.}~\bibnamefont{Moretti}}, \bibnamefont{and}
  \bibinfo{author}{\bibfnamefont{B.~R.} \bibnamefont{Webber}},
  \emph{\bibinfo{title}{{Better Jet Clustering Algorithms}}},
  \bibinfo{journal}{JHEP} \textbf{\bibinfo{volume}{08}}, \bibinfo{pages}{001}
  (\bibinfo{year}{1997}), \eprint{hep-ph/9707323}.

\bibitem[{\citenamefont{Wobisch and Wengler}(1998)}]{Wobisch:1998wt}
\bibinfo{author}{\bibfnamefont{M.}~\bibnamefont{Wobisch}} \bibnamefont{and}
  \bibinfo{author}{\bibfnamefont{T.}~\bibnamefont{Wengler}},
  \emph{\bibinfo{title}{{Hadronization Corrections to Jet Cross Sections in
  Deep- Inelastic Scattering}}} (\bibinfo{year}{1998}),
  \eprint{hep-ph/9907280}.

\bibitem[{\citenamefont{Alwall et~al.}(2007)}]{Alwall:2007st}
\bibinfo{author}{\bibfnamefont{J.}~\bibnamefont{Alwall}} \bibnamefont{et~al.},
  \emph{\bibinfo{title}{{MadGraph/MadEvent v4: The New Web Generation}}},
  \bibinfo{journal}{JHEP} \textbf{\bibinfo{volume}{09}}, \bibinfo{pages}{028}
  (\bibinfo{year}{2007}), \eprint{0706.2334}.

\bibitem[{\citenamefont{Sjostrand et~al.}(2006)\citenamefont{Sjostrand, Mrenna,
  and Skands}}]{pythiamanual}
\bibinfo{author}{\bibfnamefont{T.}~\bibnamefont{Sjostrand}},
  \bibinfo{author}{\bibfnamefont{S.}~\bibnamefont{Mrenna}}, \bibnamefont{and}
  \bibinfo{author}{\bibfnamefont{P.}~\bibnamefont{Skands}},
  \emph{\bibinfo{title}{{PYTHIA 6.4 Physics and Manual}}},
  \bibinfo{journal}{JHEP} \textbf{\bibinfo{volume}{05}}, \bibinfo{pages}{026}
  (\bibinfo{year}{2006}), \eprint{hep-ph/0603175}.

\bibitem[{\citenamefont{Cacciari and Salam}(2006)}]{Cacciari:2005hq}
\bibinfo{author}{\bibfnamefont{M.}~\bibnamefont{Cacciari}} \bibnamefont{and}
  \bibinfo{author}{\bibfnamefont{G.~P.} \bibnamefont{Salam}},
  \emph{\bibinfo{title}{{Dispelling the $N^{3}$ Myth for the $k_t$
  Jet-Finder}}}, \bibinfo{journal}{Phys. Lett.}
  \textbf{\bibinfo{volume}{B641}}, \bibinfo{pages}{57} (\bibinfo{year}{2006}),
  \eprint{hep-ph/0512210}.

\bibitem[{\citenamefont{{ The CMS Collaboration}}(2006)}]{CMSTDR}
\bibinfo{author}{\bibnamefont{{ The CMS Collaboration}}},
  \emph{\bibinfo{title}{{CMS Technical Design Report}}} (\bibinfo{year}{2006}).

\bibitem[{\citenamefont{Ellis et~al.}(1988)\citenamefont{Ellis, Hinchliffe,
  Soldate, and van~der Bij}}]{Ellis:1987xu}
\bibinfo{author}{\bibfnamefont{R.~K.} \bibnamefont{Ellis}},
  \bibinfo{author}{\bibfnamefont{I.}~\bibnamefont{Hinchliffe}},
  \bibinfo{author}{\bibfnamefont{M.}~\bibnamefont{Soldate}}, \bibnamefont{and}
  \bibinfo{author}{\bibfnamefont{J.~J.} \bibnamefont{van~der Bij}},
  \emph{\bibinfo{title}{{Higgs Decay to $\tau^+ \tau^-$: A Possible Signature
  of Intermediate Mass Higgs Bosons at the SSC}}}, \bibinfo{journal}{Nucl.
  Phys.} \textbf{\bibinfo{volume}{B297}}, \bibinfo{pages}{221}
  (\bibinfo{year}{1988}).

\bibitem[{\citenamefont{Rainwater et~al.}(1999)\citenamefont{Rainwater,
  Zeppenfeld, and Hagiwara}}]{Rainwater:1998kj}
\bibinfo{author}{\bibfnamefont{D.~L.} \bibnamefont{Rainwater}},
  \bibinfo{author}{\bibfnamefont{D.}~\bibnamefont{Zeppenfeld}},
  \bibnamefont{and} \bibinfo{author}{\bibfnamefont{K.}~\bibnamefont{Hagiwara}},
  \emph{\bibinfo{title}{{Searching for $H \to \tau \tau$ in Weak Boson Fusion
  at the LHC}}}, \bibinfo{journal}{Phys. Rev.} \textbf{\bibinfo{volume}{D59}},
  \bibinfo{pages}{014037} (\bibinfo{year}{1999}), \eprint{hep-ph/9808468}.

\bibitem[{\citenamefont{Plehn et~al.}(2000)\citenamefont{Plehn, Rainwater, and
  Zeppenfeld}}]{Plehn:1999xi}
\bibinfo{author}{\bibfnamefont{T.}~\bibnamefont{Plehn}},
  \bibinfo{author}{\bibfnamefont{D.~L.} \bibnamefont{Rainwater}},
  \bibnamefont{and}
  \bibinfo{author}{\bibfnamefont{D.}~\bibnamefont{Zeppenfeld}},
  \emph{\bibinfo{title}{{A Method for Identifying $H \to \tau \tau \to e^{+-}
  \mu^{-+}$ Missing p(T) at the CERN LHC}}}, \bibinfo{journal}{Phys. Rev.}
  \textbf{\bibinfo{volume}{D61}}, \bibinfo{pages}{093005}
  (\bibinfo{year}{2000}), \eprint{hep-ph/9911385}.

\bibitem[{\citenamefont{Aad et~al.}(2009)}]{Aad:2009wy}
\bibinfo{author}{\bibfnamefont{G.}~\bibnamefont{Aad}} \bibnamefont{et~al.}
  (\bibinfo{collaboration}{The ATLAS}), \emph{\bibinfo{title}{{Expected
  Performance of the ATLAS Experiment - Detector, Trigger and Physics}}}
  (\bibinfo{year}{2009}), \eprint{0901.0512}.

\end{thebibliography}
\bibliographystyle{apsper}

\end{document}